\documentclass[usenatbib]{mnras}
\usepackage{lscape}
\usepackage{amsmath,amssymb}
\usepackage{mathrsfs}	
\usepackage{multicol}
\usepackage{textcomp}
\usepackage[dvipdfmx]{graphicx}
\usepackage{xcolor}

\graphicspath{{figures/}}

\catcode`\@=11
\def\gta{\ifmmode{\,\mathrel{\mathpalette\@versim>\,}}
    \else{$\,\mathrel{\mathpalette\@versim>}\,$}\fi}
\def\lta{\ifmmode{\,\mathrel{\mathpalette\@versim<\,}}
    \else{$\,\mathrel{\mathpalette\@versim<}\,$}\fi}
\def\@versim#1#2{\lower 2.9truept \vbox{\baselineskip 0pt \lineskip
    0.5truept \ialign{$\m@th#1\hfil##\hfil$\crcr#2\crcr\sim\crcr}}}
\catcode`\@=12  

{\newif\ifnotend
\notendtrue
\def\veclist{ABCDEFGHIJKLMNOPQRSTUVWXYZabcdefghijklmnopqrstuvwxyz.}
\def\top#1#2.{#1}
\def\tail#1#2.{#2.}
\loop\expandafter\xdef\csname bb\expandafter\top\veclist\endcsname%
{{\noexpand\bf\expandafter\top\veclist}}
\edef\veclist{\expandafter\tail\veclist}
\if\veclist.\notendfalse\fi\ifnotend\repeat}
\newcommand{\bbsigma}{\boldsymbol \sigma}

\makeatletter
\def\blfootnote{\xdef\@thefnmark{}\@footnotetext}
\makeatother

\newcommand{\fxv}	    {f({\bf x},{\bf v})}
\newcommand{\fnorm}	    {f_0}

\newcommand{\fEnorm}	{f_{0,E}}
\newcommand{\fLz}   	{f({L_z})}
\newcommand{\fLznorm}   {f_{0,L_z}}
\newcommand{\Rc}	    {R_{\rm c}}
\newcommand{\too}       {t_0}
\newcommand{\Ro}        {R_0}
\newcommand{\Reff}      {R_{\rm eff}}
\newcommand{\rh}        {r_{\rm h}}
\newcommand{\rhonorm}	{\rho_0}
\newcommand{\Lz}	    {L_z}
\newcommand{\Phieff}	{\Phi_{\rm eff}}

\newcommand{\mmax}	    {m_{\rm max}}

\newcommand{\Rizi}	{\{R_i,z_i\}}

\newcommand{\vx}	{v_x}
\newcommand{\vy}	{v_y}
\newcommand{\vz}	{v_z}
\newcommand{\xk}	{x_k}
\newcommand{\yk}	{y_k}
\newcommand{\zk}	{z_k}
\newcommand{\vxk}	{v_{x,k}}
\newcommand{\vyk}	{v_{y,k}}
\newcommand{\vzk}	{v_{z,k}}

\newcommand{\vR}	{v_R}
\newcommand{\vphi}	{v_\phi}
\newcommand{\pR}	{p_R}
\newcommand{\pz}	{p_z}
\newcommand{\pphi}	{p_\phi}

\newcommand{\tr}	{\tilde{r}}
\newcommand{\tm}	{\widetilde{m}}
\newcommand{\tmmax}	{\widetilde{m}_{\rm max}}

\newcommand{\Mc}	{M_{\rm c}}
\newcommand{\Mt}	{M_{\rm t}}
\newcommand{\Mmin}	{M_{\rm min}}
\newcommand{\multiN}	{\mathcal{M}}

\newcommand{\Ltot}	{\mathcal{L}}

\newcommand{\Nc}	{N_{\rm c}}
\newcommand{\Nt}	{N_{\rm t}}
\newcommand{\Nu}	{N_{\rm u}}
\newcommand{\Nr}	{N_{\rm r}}

\newcommand{\xir}	{\xi_{\rm r}}
\newcommand{\xic}	{\xi_{\rm c}}
\newcommand{\xit}	{\xi_{\rm t}}
\newcommand{\xiu}	{\xi_{\rm u}}
\newcommand{\BB}	{\mathcal{B}}

\newcommand{\dd}    	{\text{d}}
\newcommand{\DD}	    {\partial}
\newcommand{\Vc}	{{\mathcal V}_{\rm c}}
\newcommand{\Vt}	    {{\mathcal V}_{\rm t}}
\newcommand{\Nchain}    {N_{\rm chain}}
\newcommand{\Nbi}       {N_{\rm burn-in}}

\newcommand{\Lcirc}    {L_{\rm circ}}
\newcommand{\barR}  {\bar{R}}

\newcommand{\rapex}	{\textacutedbl}
\newcommand{\lapex}	{\textgravedbl}

\newcommand{\Rtilde}    {\tilde{R}}
\newcommand{\ztilde}    {\tilde{z}}
\newcommand{\ta}        {\tilde{a}}
\newcommand{\tz}        {\tilde{z}}
\newcommand{\tR}        {\tilde{R}}
\newcommand{\epsz}      {\varepsilon_z}

\newcommand{\Stackel}   {St\"ackel\,}
\newcommand{\kpc}	{\, {\rm kpc}}
\newcommand{\Myr}   {\, {\rm Myr}}

\newcommand{\kms}	{\, {\rm km \,s^{-1}}}

\newcommand{\imaginary}{{\rm i}}
\renewcommand{\d}{{\rm d}}

\definecolor{darkgreen}{rgb}{0.0, 0.3, 0.0}

\begin{document}

\date{}
\title[Regular and chaotic orbits in stellar systems]
{Regular and chaotic orbits in axisymmetric stellar systems}
{}
\author[]{Raffaele Pascale$^{1,2}$\thanks{E-mail: raffaele.pascale@inaf.it}, Carlo Nipoti$^{2}$ and Luca Ciotti$^{2}$
\\ \\
$^{1}$INAF - Osservatorio di Astrofisica e Scienza dello Spazio di Bologna, via Piero Gobetti 93/3, I-40129 Bologna, Italy\\
$^{2}$Dipartimento di Fisica e Astronomia \lapex Augusto Righi\rapex, Universit\`a di Bologna, via Piero Gobetti 93/2, I-40129 Bologna, Italy}

\maketitle

\begin{abstract}
The gravitational potentials of realistic galaxy models are in general non-integrable, in the sense that they admit orbits that do not have three independent isolating integrals of motion and are therefore chaotic. However, if chaotic orbits are a small minority in a stellar system, it is expected that they have negligible impact on the main dynamical properties of the system. In this paper we address the question of quantifying the importance of chaotic orbits in a stellar system, focusing, for simplicity, on axisymmetric systems. Chaotic orbits have been found in essentially all (non-St\"ackel) axisymmetric gravitational potentials in which they have been looked for. Based on the analysis of the surfaces of section, we add new examples to those in the literature, finding chaotic orbits, as well as resonantly trapped orbits among regular orbits, in Miyamoto-Nagai, flattened logarithmic and shifted Plummer axisymmetric potentials. We define the fractional contributions in mass of chaotic ($\xic$) and resonantly trapped ($\xit$) orbits to a stellar system of given distribution function, which are very useful quantities, for instance in the study of the dispersal of stellar streams of galaxy satellites.  As a case study, we measure $\xic$ and $\xit$ in two axisymmetric stellar systems obtained by populating flattened logarithmic potentials with the Evans ergodic distribution function, finding $\xic\sim 10^{-4}-10^{-3}$ and $\xit\sim 10^{-2}-10^{-1}$.
\end{abstract}

\begin{keywords}
celestial mechanics -- chaos -- galaxies: kinematics and dynamics -- methods: numerical -- methods: statistical
\end{keywords}

\section{Introduction}
\label{sec:intro}

In stellar dynamics, the most important class of axisymmetric and triaxial gravitational potentials known to be integrable are those of the \cite{Stackel1893} family: the separability of the Hamilton-Jacobi equation in ellipsoidal coordinates ensures the existence of three global independent isolating integrals of motion and thus that all orbits are regular (see e.g.\ \citealt{deZeeuw1985} and \citealt{deZeeuwLyndenBell1985}).
The gravitational potentials of realistic galactic models, which in general are not of \Stackel form, are not guaranteed to be integrable and can admit chaotic orbits, i.e.\ orbits that do not have three independent isolating integrals of motion. The integrability of non-spherical galactic potentials and the 
contribution of chaotic orbits to stellar systems with non-integrable potentials are important questions for galactic dynamics, which can be addressed by classifying samples of numerically integrated orbits. 

In the present work we focus on axisymmetric potentials. Though chaos is often studied in the context of triaxial potentials \citep[e.g.][]{Schwarzschild1979,Schwarzschild1982,MiraldaEscude1989,Valluri1998}, chaotic orbits are found also in axisymmetric potentials (e.g. \citealt{HenonHeiles1964,Hunter2003,Hunter2005,Zotos2013}). In fact, chaotic orbits have been found in essentially all (non-St\"ackel) axisymmetric potentials in which they have been looked for. \citet{Hunter2005} suggested that a possible exception could be the family of Miyamoto-Nagai (hereafter MN) potentials \citep{MiyamotoNagai1975}, whose limiting cases---the \citet{Plummer1911} sphere and the \citet{Kuzmin1956} disc---are integrable, but it turns out this is not the case, because, as we show in this paper, the MN family admits chaotic orbits. Using the surface of section (SoS) method \citep{HenonHeiles1964,BinneyTremaine2008}, we add new examples to those previously considered in the literature, demonstrating that a few more families of axisymmetric potentials are non-integrable (at least for some values of their parameters): not only the aforementioned MN potential, but also the shifted Plummer axisymmetric and the flattened logarithmic potentials.


Of course, while it is sufficient to find a chaotic orbit to demonstrate that a given potential is non-integrable, it is impossible to proof with numerical integration of orbits that a given potential is integrable. However, for many practical purposes it is not so important to determine whether a given potential is integrable or not, but rather to estimate the fractional contribution of chaotic orbits to a stellar system. There is general consensus, mainly based on the results of numerical investigations \citep{Schwarzschild1979,Richstone1982}, that in realistic galactic potentials \lapex most of the orbits are regular\rapex and that reasonable dynamical models can be built by neglecting the small fraction of chaotic orbits. However, especially in an era in which the galactic dynamical models are more and more sophisticated  \citep[e.g.][]{Binney2020} and the observational data are characterised by high resolution and high statistics \citep{Rave2013,Apogee2017,Gaia2018,Gaia2018b,Gaia2020}, one would really like to make more quantitative statements about the relative contributions of regular and chaotic orbits. An astrophysical application in which the knowledge of these contributions is highly relevant is the study of stellar streams
of galaxy satellites, often used as tracers of the host galaxy gravitational field
and thus mass density distribution \citep{Helmi1999,Fardal2015,Bonaca2020,Mestre2020}. In essence, the time over which a stream is dispersed is expected to be much shorter if the orbit of its progenitor satellite (globular cluster of dwarf galaxy) is chaotic than if it is regular. In some cases even orbits with extremely long characteristic chaotic timescales can produce chaotic effects on relatively short timescales \citep{PriceWhelan2016}.

Quantitative estimates of the fractional contributions of chaotic and regular orbits to stellar systems are rare in the literature. \citet{Maffione2015}, extracting particles from solar neighbourhood-like volumes of dark-matter only cosmological simulations of Milky-Way like halos, estimated the fractions of chaotic and regular orbits, integrated in triaxial analytic approximations of the simulated gravitational potentials. A similar systematic study was carried out by \citet{Maffione2018}, who considered more realistic $N$-body realizations of Milky-Way like galaxies extracted from hydrodynamic cosmological simulations. Here, adopting a non-cosmological, but fully self-consistent approach, we consider stellar systems with given distribution function (DF), for which we provide  rigorous definitions of the fractional contributions in mass of regular and chaotic orbits.

Regular orbits include the special family of resonantly trapped orbits. While a resonant orbit is such that its fundamental frequencies  are commensurable, an orbit is said to be {\em resonantly trapped} when it is, in phase space, close to a resonant orbit (see \citealt{BinneyTremaine2008}). Resonantly trapped orbits must be treated with special care, since they behave differently from the other regular orbits in the angle-action space \citep{Binney2016}. Also the aforementioned dispersal of stellar streams can be significantly affected, or even dominated, by the presence of families of resonantly trapped orbits in the host gravitational field.  The time over which a stream is dispersed can be relatively short if the orbit of the progenitor satellite is close to a separatrix, i.e.\  the boundary between two orbit families, each defined by a different orbital resonance (a phenomenon known as separatrix divergence; \citealt{Yavetz2021}). In the following we will refer to regular orbits that are not resonantly trapped as {\em untrapped}. Thus, in addition to chaotic orbits, in this work we classify separately also resonantly trapped and untrapped regular orbits, detecting them with the SoS method and measuring their fractional mass contributions to a stellar system.

As a case study, we present the results obtained by estimating the fractional contributions in mass of regular and chaotic orbits in flattened axisymmetric stellar systems with luminous component with \citet{Evans1993} DF, embedded in an external logarithmic gravitational potential. Evans' models, though idealized in some respects (they are isotropic and they do not account explicitly for the self-gravity of the luminous component), are interesting for applications in galactic dynamics, because their luminous density distribution is, for a range of values of their free parameters, a reasonably good model of the luminous density distribution of galactic spheroids. The flattened axisymmetric logarithmic potential has been widely used in applications to galactic dynamics: gravitational potentials belonging to this family has been adopted, for instance, by \citet{Helmi2004a,Helmi2004b} to model the dark matter halo of the Milky Way, by \citet{Bovy2014} to study tidal streams, and by \citet{Sanders2016} to model dark matter halos of dwarf spheroidal galaxies.  More recently, \citet{Hagen2019} used Evans models to build mock dwarf spheroidal galaxy models.

The paper is organised as follows. After recalling the definition of the SoS in Section~\ref{sec:sos} and describing the adopted numerical methods in Section~\ref{sec:numerical},  in Section~\ref{sec:orbclass} we present new results on SoS-based classification of orbits in axisymmetric potentials. In Section~\ref{sec:chaos} we define the fractional mass contributions of regular and chaotic orbits to a stellar system with given DF, and we present the results of our case study. Section~\ref{sec:conc} concludes.

\begin{figure*}
    \centering
    \includegraphics[width=1\hsize]{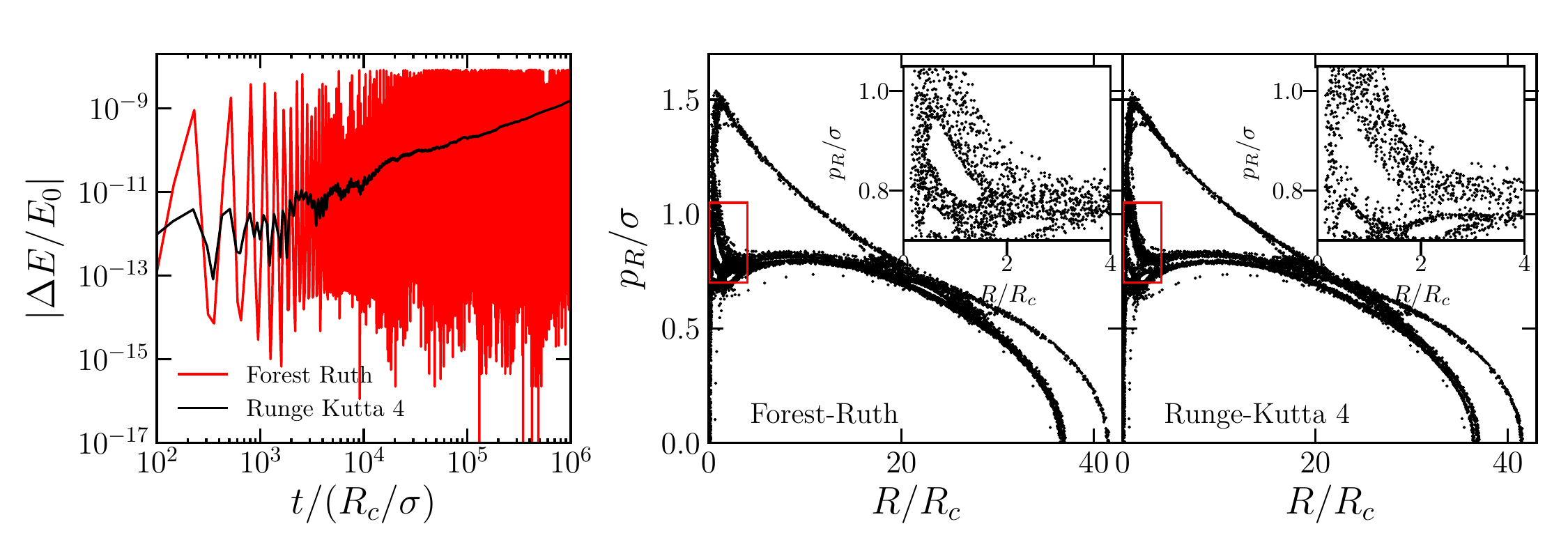}
    \caption{Left panel: relative energy variation as a function of time for a chaotic orbit integrated in the potential~(\ref{for:pot}) with $q=0.75$ and $\Rc=\sigma=1$, using the symplectic Forest-Ruth algorithm (red curve) and the RK 4 scheme (black curve) described in Section~\ref{sec:integrator}. Middle panel: trace in the SoS of the same orbit as in the left panel integrated with the Forest-Ruth scheme. Right panel: same as the middle panel, but using the adaptive RK 4 scheme. The small insets in the middle and right panels show zoomed-in views of the red boxes in the corresponding main panels.}
    \label{fig:econs}
\end{figure*}


\section{Surfaces of section for orbits in axisymmetric potentials}
\label{sec:sos}

Let $\Phi(R,z)$ be an axisymmetric potential in cylindrical coordinates $(R,\phi,z)$, and let $\gamma$ be the trajectory in phase space of a star moving under the action of $\Phi$. The Poincaré map or SoS is the plane having ($R,\pR$) as axes, where $\pR$ is the momentum conjugated to $R$. The trace of an orbit in the SoS is the set of the  orbit's consequents, i.e.\  the points with coordinates ($R,\pR$) defined by the intersections of $\gamma$ with the equatorial plane ($z=0$). Potentials that are time-independent and invariant with respect to rotation around a given axis (chosen to be the $z$-direction) admit as global integrals of motion the specific (i.e.\ per unit mass) energy $E$ and the $z$-component of the specific angular momentum $\Lz$. For given $E$ and $\Lz$, the points of a trace belong, by construction, to the manifold defined by the implicit function

\begin{equation}\label{for:manifold}
 E = \frac{1}{2}(\pR^2 + \pz^2)+\Phieff(R,0),
\end{equation}
where $\pz$ is the momentum conjugated to $z$ and
\begin{equation}\label{for:phieff}
 \Phieff(R,z)=\frac{\Lz^2}{2R^2} + \Phi(R,z)
\end{equation}
is the effective potential.

While the manifold (\ref{for:manifold}) is two-dimensional, the trace of an orbit in the SoS can be either one- or two-dimensional. For given $E$ and $\Lz$, the consequents of an orbit are confined to a region which is given by the relation
\begin{equation}
 E - \Phieff(R,0) \geq 0.
\end{equation}
Also, traces of different orbits with the same values of $E$ and $\Lz$ cannot cross each other in the SoS: if the traces of two orbits intersected at one or more points in the SoS, there would exist a point on $\gamma$ where the two orbits have the same $(E,\Lz,R,\pR)$ at $z=0$. Since equation~(\ref{for:manifold}) implies that the two orbits can only differ in the sign of $\pz$, so they have the same trajectory. The sign ambiguity is usually resolved by tracing a point on the SoS only when the crossing of the equatorial plane occurs with $\pz>0$. Here we do not conform to this convention but, as done for instance in \citet{Richstone1982} and \citet{Hunter2005}, we trace a point in the SoS for crossings with both $\pz>0$ and $\pz<0$. We find this choice convenient because, following \citet{Richstone1982}, we then classify an orbit as resonantly trapped if its trace in the SoS consists of two or more unconnected loops.


For axially symmetric potentials, given a trace of consequents, the existence of a third isolating integral is deduced by the dimensionality of the trace in the SoS. A one-dimensional trace implies that the motion is further constrained onto an additional surface, defined, for instance, by $\pR=\pR(R,z,\pz)$, leading to the conclusion that the orbit is regular. Instead, when there is no additional integral of motion, the consequents populate two-dimensional regions of the SoS and the orbit is chaotic. SoS have been used to study the existence of a third isolating integral since the mid 1960's \citep[e.g.][]{HenonHeiles1964,Bienayme2015}.

\section{Numerical methods}
\label{sec:numerical}

Before presenting the results obtained applying the SoS method to classify orbits in a few axisymmetric potentials (Sections~\ref{sec:orbclass} and~\ref{sec:chaos}),  we describe here the adopted numerical tools that allow us to obtain the required accuracy in the computation of the orbits' traces in the SoS.

\subsection{Numerical integration of the orbits }
\label{sec:integrator}

As done in other works on orbit classification in galactic potentials \citep[e.g.][]{Richstone1982,PriceWhelan2016}, we opted for a Runge-Kutta (RK) algorithm to integrate numerically the equations of motion. Specifically, we computed the orbits using an adaptive fourth-order RK integrator (hereafter RK 4; \citealt{Butcher1996}). The integration is performed in Cartesian coordinates according to the following scheme (\citealt{WilliamSaul1992}):

\begin{itemize}
 \item[i)] Let $\bbw=(x,y,z,\vx,\vy,\vz)$. Starting from the phase-space position $\bbw_n$ at a time $t_n$, we evaluate the proposals $\bbw_{n+1}^{\Delta t_n}$ and $\bbw_{n+1}^{\Delta t_n/2}$ at a subsequent time $t_{n+1}$, corresponding, respectively, to time steps $\Delta t_n$ and $\Delta t_n/2$. 

 \item[ii)] We evaluate $\epsilon_n = \max_i|1-w_{n+1,i}^{\Delta t_n}/w_{n+1,i}^{\Delta t_n/2}|$ and compare it to some predetermined accuracy $\epsilon$. $w_{n+1,i}^{\Delta t_n}$ and $w_{n+1,i}^{\Delta t_n/2}$, with $i=1,...,6$, are the $i$-th elements of the vectors $\bbw_{n+1}^{\Delta t_n}$ and $\bbw_{n+1}^{\Delta t_n/2}$, respectively.
 
 \item[iii.a)] If $\epsilon_n>\epsilon$, both proposals $\bbw_{n+1}^{\Delta t_n}$ and $\bbw_{n+1}^{\Delta t_n/2}$ are rejected, and new proposals for $\bbw_{n+1}^{\Delta t_n}$ and $\bbw_{n+1}^{\Delta t_n/2}$ are found using as new time step 
 \begin{equation}\label{for:RKt1}
  \Delta t_n \beta (\epsilon/\epsilon_n)^{1/(h+1)},
 \end{equation}
 and its half, respectively, where $\beta$ and $\epsilon$ are dimensionless parameters, and $h=4$ is the order of the Runge-Kutta scheme.
 \item[iii.b)] If $\epsilon_n<\epsilon$, $\bbw_{n+1}=\bbw_{n+1}^{\Delta t_n/2}$ and the new timestep is
 \begin{equation}\label{for:RKt2}
  \Delta t_n \beta (\epsilon/\epsilon_n)^{1/h}.
 \end{equation}
\end{itemize}

\begin{figure*}
    \centering
    \includegraphics[width=1\hsize]{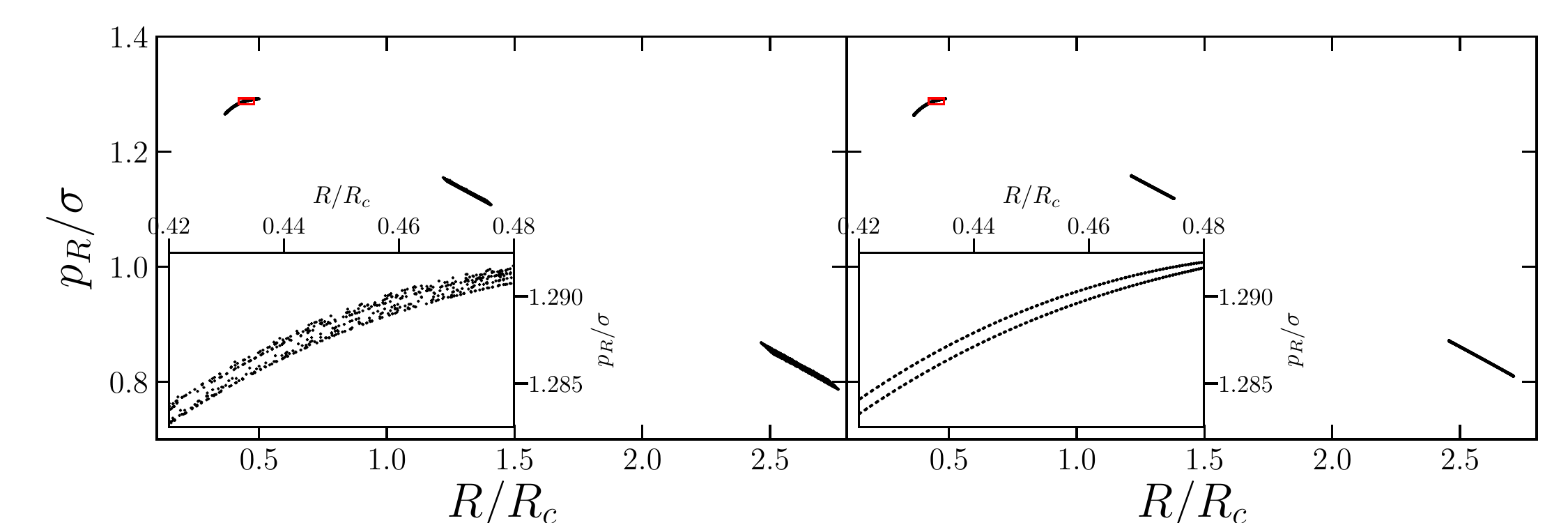}
    \caption{Left panel: trace in the SoS of a test resonantly trapped orbit integrated in the potential~(\ref{for:pot}) whose crossings of the equatorial plane have been evaluated using $\epsz=10^{-1}$. Right panel: same as the left panel, but with $\epsz=10^{-6}$.}
    \label{fig:cross}
\end{figure*}

RK algorithms are known to be dissipative, in the sense that they do not ensure the conservation of the mechanical energy in a Hamiltonian system. As shown for instance by \citet{Stuchi2002}, if the numerical integration of a test particle in a Hamiltonian system proceeds for sufficiently long time, when using dissipative algorithms the energy drifts because of numerical dissipation and the consequent deviation of the integrated phase-space position from the true one can in principle shift the target particle from a region of regularity to a region of chaos. This can make an intrinsically regular orbit appear as chaotic after a long-time integration.
For this reason, in the context of numerical integration of orbits and their classification as regular or chaotic, it is sometimes preferred the use of symplectic integrators, which are not dissipative \citep[e.g.][]{Barnes2001,Mestre2020}. 

Our adaptive RK scheme implementation allows us to keep very high precision in the orbit integration and minimize numerical dissipation. In equations (\ref{for:RKt1}) and (\ref{for:RKt2}) we set $\beta=0.9$ and $\epsilon=10^{-12}$, obtaining energy conservation of a part over $10^{12}-10^9$, depending on the maximum time integration required, which is a strong indication that the effects of dissipation should be negligible. As an additional check that our results are not significantly affected by numerical effects related to accuracy and the dissipative nature of the adopted integrator, we integrated all the orbits classified as chaotic based on the RK 4 integration with $\epsilon=10^{-12}$, also requiring $\epsilon=10^{-15}$ in equations~(\ref{for:RKt1}) and (\ref{for:RKt2}), and also with the 4-th order symplectic Forest-Ruth integrator \citep{Forest1990,Yoshida1990,Candy1991}, finding very good agreement between the numerical results and confirming the classification.
In Fig.~\ref{fig:econs} we compare the relative energy of a chaotic orbit integrated in the potential~(\ref{for:pot}) using our implementation of the RK 4 scheme and the aforementioned symplectic Forest-Ruth scheme. Although the RK scheme introduces numerical dissipation, the energy conservation is always comparable with the energy oscillations produced by the Forest-Ruth scheme and, as shown in the middle- and right-hand panels of Fig.~\ref{fig:econs}, the orbit manifests its chaotic behavior when integrated with both algorithms. In terms of time performance, our implementation of the adaptive RK 4 scheme is 1.5-7 times faster than the considered Forest-Ruth scheme (depending on the specific orbit) and allows us to evaluate very precisely the phase-space coordinates at the times of crossing of the equatorial plane (see Section~\ref{sec:traces}).

\begin{figure*}
 \centering
 \includegraphics[width=.9\hsize]{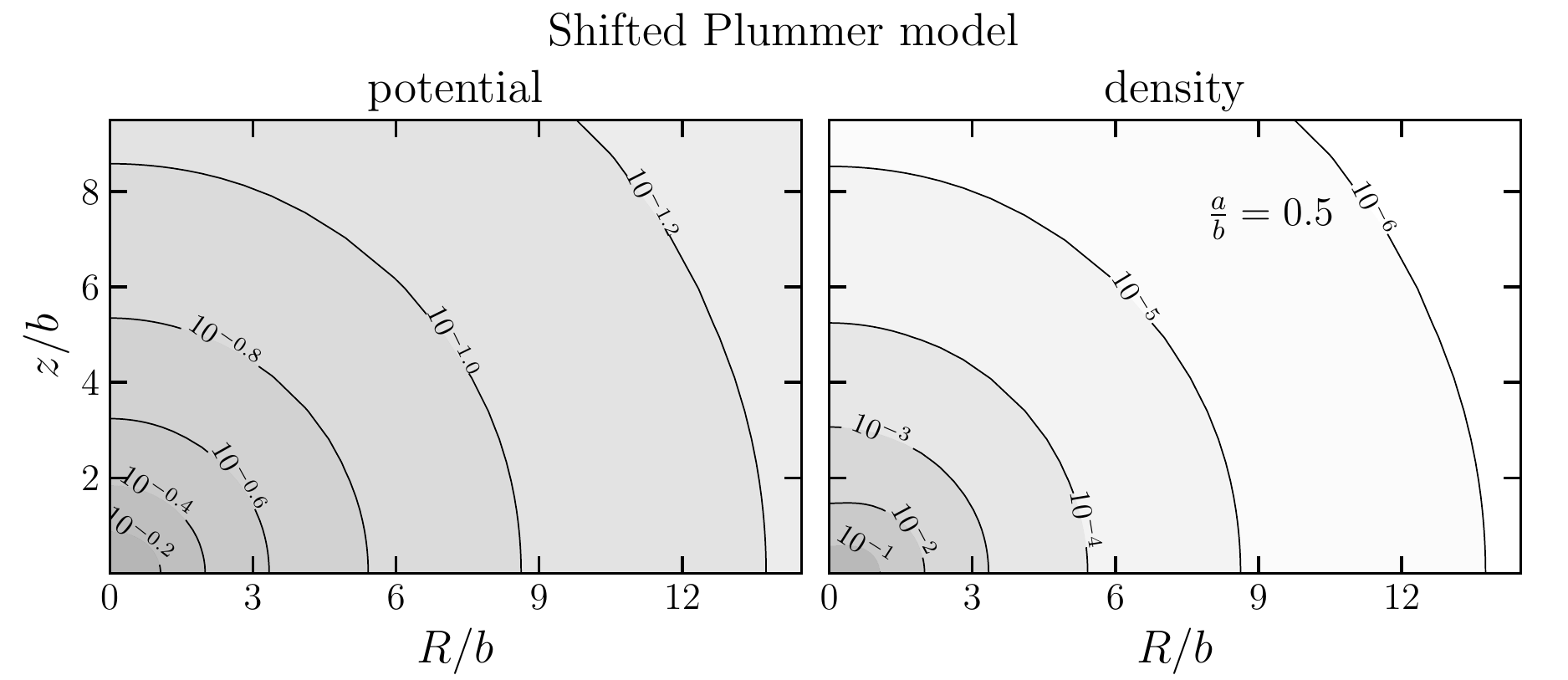}
 \caption{Left panel: iso-potential contours in the $(R,z)$-plane of the shifted Plummer model (equation \ref{for:shiftPlum}) with $a/b=0.5$, labelled by the values of $\Phi/\Phi_0$, where $\Phi_0$ is the central potential. Right panel: iso-density contours of the same model as in the left panel, labelled by the values of $\rho/\rho_0$, where $\rho_0$ is the central density.}
 \label{fig:SP}
\end{figure*}

\begin{table*}
\centering
 \begin{tabular}{lcccccc}
 \hline\hline \multicolumn{7}{c}{Shifted Plummer potential} \\ \hline\hline
    orbit   & $R/b$   & $z/b$ &   $\pR/\sqrt{GM/b}$ & $\pz/\sqrt{GM/b}$   &   $E/(GM/b)$	&   $\Lz/\sqrt{GMb}$    \\
 \hline\hline   
    regular & 0.4   &  	0       &   0   &   1.413       &   -0.05   &   $10^{-3}$   \\
    chaotic & 0.35  &   0       &   0   &   1.42869     &   -0.05   &   $10^{-3}$   \\
    trapped & 0.28  &   0    &   0.18   &   1.43701     &   -0.05   &   $10^{-3}$  \\
   \hline\hline
 \end{tabular}
 \caption{Initial conditions in cylindrical coordinates, specific energy $E$ and vertical component of the specific angular momentum $\Lz$ of the regular, chaotic and resonantly trapped orbits of Fig.~\ref{fig:SPsos} ($\phi$ is arbitrary and $\pphi=\Lz$). The orbits are integrated in the shifted Plummer potential (\ref{for:shiftPlum}) with $a/b=0.5$.}
 \label{tab:icSP}
\end{table*}

\begin{figure*}
 \centering
 \includegraphics[width=1\hsize]{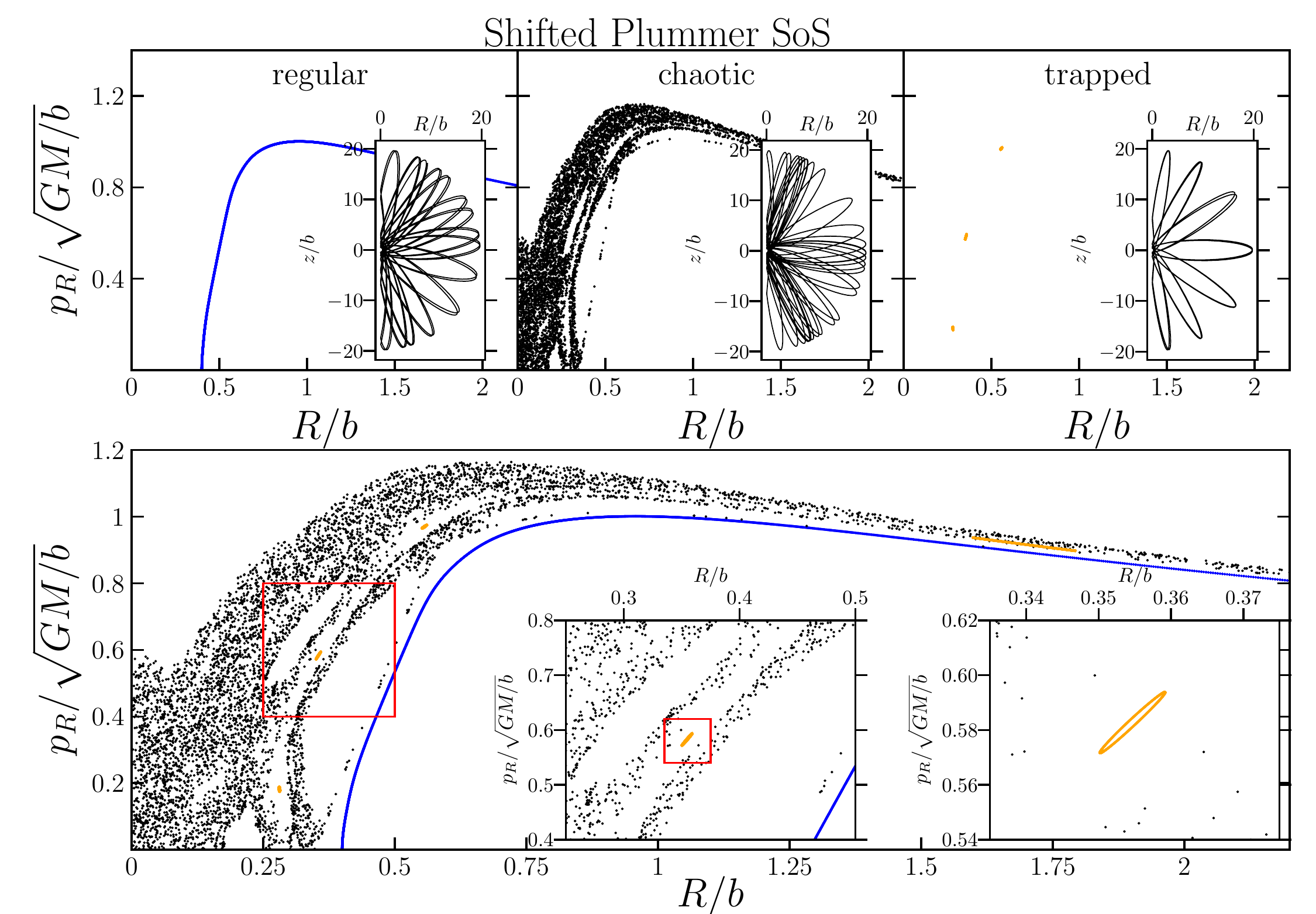}
 \caption{Top panels: traces in the SoS of a regular orbit (blue points, left panel), a chaotic orbit (black points, middle panel) and a resonantly trapped orbit (orange points, right panel) in the shifted Plummer potential (\ref{for:shiftPlum}) with $a/b=0.5$. In each panel the inset shows the trajectory of the corresponding orbit in the meridional plane. Bottom panel: traces of the same regular, chaotic and resonantly trapped orbits as in the top panels, but plotted all in the same SoS. The left inset shows a zoomed-in view of the region marked with a red box in the main panel. The right inset shows a zoomed-in view of the region marked with a red box in the left inset. The ICs of the orbits are given in Table \ref{tab:icSP}.}\label{fig:SPsos}
\end{figure*}

\subsection{Computation of the traces in the surface of section}
\label{sec:traces}

Central to evaluate consequents in the SoS is the ability to compute as precisely as possible $\bbw$ at the time corresponding to each crossing of the equatorial plane. Each time $z_n z_{n+1}<0$ (i.e.\ a crossing through the equatorial plane has occurred), together with equations (\ref{for:RKt1}) and (\ref{for:RKt2}), we make the further requirement that $|z_{n+1}-z_n|/R_0 < \epsz$, where $\Ro$ is a characteristic radius, for instance $\Ro\equiv\Rc$ in case of the logarithmic potential of Section~\ref{sec:casestudy}. When the condition is not satisfied a new guess for $z_{n+1}$ is made halving the time step. Such condition ensures that $z_{n+1}$ is close to the equatorial plane to a precision dictated by $\epsz$. Throughout this work, we have adopted $\epsz=10^{-6}$. In Fig.~\ref{fig:cross} we show the consequents in the SoS of an orbit integrated into potential~(\ref{for:pot}) using $\epsz=10^{-1}$ (left panel) and $\epsz=10^{-6}$ (right panel). The traces belong to a regular trapped orbit (see also Section~\ref{sec:plummer}), whose loops are so tight in the SoS that, if the crossings of the equatorial plane are not precisely evaluated (as in the left panel), the orbit could be misclassified as chaotic. The inset in the right panel of Fig.~\ref{fig:cross} demonstrates that the choice $\epsz=10^{-6}$ guarantees the required accuracy.

\section{Finding chaotic orbits in axisymmetric gravitational potentials}
\label{sec:orbclass}

Here we apply the SoS method to classify orbits in two families of axisymmetric potentials: the shifted Plummer and MN potentials. Orbits in the shifted Plummer potentials, as far as we are aware, have not been studied before. Orbits in a MN potential have been classified by \citet{Hunter2005} and \citet{Zotos2013}, who found only regular orbits \citep[see also][]{Greiner1987,Greiner1990}. In Sections~\ref{sec:plummer} and \ref{sec:mn} we present examples of potentials belonging to these families, showing that they admit, together with regular orbits, also chaotic orbits. Among regular orbits, we identify also members of the special family of resonantly trapped orbits. A third family of gravitational potentials, the flattened axisymmetric logarithmic potentials \citep{Binney1981}, is considered in Section~\ref{sec:casestudy}. Orbits in these logarithmic potentials have been studied in previous works \citep{Richstone1982,Barnes2001}, but, to our knowledge, only regular orbits have been found so far: in Section~\ref{sec:casestudy}  we show examples of chaotic orbits in two potentials of this family.

\subsection{Orbits in a shifted Plummer potential}
\label{sec:plummer}

The complexification is a shift $\bbx\to\bbx-\imaginary\bba$ that maps a potential $\Phi(\bbx)$ into $\Phi(\bbx-\imaginary\bba)$, where $\bba=(a_1,a_2,a_3)$ is a triplet of real numbers. As shown by \cite{CiottiGiampier2007} and \cite{CiottiMarinacci2008}, the complex shift method can be used to obtain analytic density-potential pairs $(\rho,\Phi)$ for axisymmetric models starting from spherical ones. If $\Phi$ satisfies the Poisson equation $\nabla^2\Phi=4\pi G\rho$ with analytic $\Phi$ and $\rho$, thanks to the linearity of the shift and of the Poisson equation, the shift gives birth to two analytic density-potential pairs given by the real and complex parts of $\Phi(\bbx-\imaginary\bba)$ and $\rho(\bbx-\imaginary\bba)$. The method was first introduced in electrostatics by \cite{Appell1887}, \cite{Whittaker1950}, \cite{Carter1968} and others, and later applied to gravitational potentials (\citealt{CiottiGiampier2007}, \citealt{CiottiMarinacci2008}, and reference therein). The complex shift method is of special interest for the purposes of this work. It is, for instance, reasonable to speculate that during the complexification of a spherical potential some of the integrability properties are transferred to the complexified versions, as shown by the surprising integrability properties of the complexified point-mass potential \citep{LyndenBell1962,LyndenBell2000,LyndenBell2003}.

Following \cite{CiottiGiampier2007}, let us consider the gravitational potential
\begin{equation}\label{for:shiftPlum}
 \Phi(R,z) = -\frac{GM}{b}\psi(R,z),
\end{equation}
where
\begin{equation}\label{for:shiftPlumpsi}
 \psi(R,z)=\sqrt{\frac{d + 1 +  \tr^2 - \ta^2}{2d^2}},
\end{equation}
with
\begin{equation}\label{for:scale}
 d = \sqrt{\left(1-\ta^2+\tr^2\right)^2 + 4\ta^2\tz^2}.
\end{equation}
Here $\tR\equiv R/b$, $\tz\equiv z/b$, $\ta\equiv a/b$ and $\tr^2=\tR^2+\tz^2$. Known as shifted Plummer model, the axisymmetric potential (\ref{for:shiftPlum}) is obtained by means of the complexification of a classical \citet{Plummer1911} sphere with mass $M$ and core radius $b$. The potential (\ref{for:shiftPlum}) corresponds to the real part of the shift of the spherical Plummer model, with $\bba=(0,0,a)$, so $a$ is the amplitude of the shift ($0\le a\le b$). The density that generates the gravitational potential (\ref{for:shiftPlum}) is 
\begin{equation}
  \begin{split}\label{for:shiftPlumdens}
 \rho = \frac{3M\psi}{4\pi b^3}\left[\psi^4 - \frac{10\ta^2\tz^2}{d^4}
 + \frac{5\ta^4\tz^4}{d^8\psi^4}\right].
  \end{split}
\end{equation}

Here we consider the shifted Plummer model with $a/b=0.5$, whose potential and density maps in the meridional plane are shown in  Fig.~\ref{fig:SP}. We have integrated numerically several orbits in this potential, finding untrapped regular orbits, chaotic orbits and resonantly trapped regular orbits. The top three panels of Fig.~\ref{fig:SPsos} show the traces in the SoS and the trajectories in the $(R,z)$-plane of three  representative orbits, having the same values of $E$ and $\Lz$ (their initial conditions are given in Table~\ref{tab:icSP}). Throughout this paper, as done for instance in \citet{Hunter2005}, we show only the $\pR>0$ part of the SoS since it is sufficient for the purpose of illustrating the nature of an orbit.
In the SoS, the consequents of the untrapped regular orbit (left panel in Fig.~\ref{fig:SPsos}) align on a one-dimensional path since, as discussed in Section \ref{sec:sos}, the orbit conserves a third isolating integral of motion, which lowers the dimensionality of the phase-space manifold on which the orbit lies. The consequents of the chaotic orbit (middle panel in Fig.~\ref{fig:SPsos}) fill a two-dimensional region, meaning that a third isolating integral of motion does not exist. The right panel of Fig.~\ref{fig:SPsos} shows the trace in the SoS of an orbit, which is regular (its trace in the SoS is one-dimensional), but trapped by resonance: the trace is a combination of circuits enclosing the points that represent, in the SoS, the parent resonant orbit.

\begin{figure*}
    \centering
    \includegraphics[width=.9\hsize]{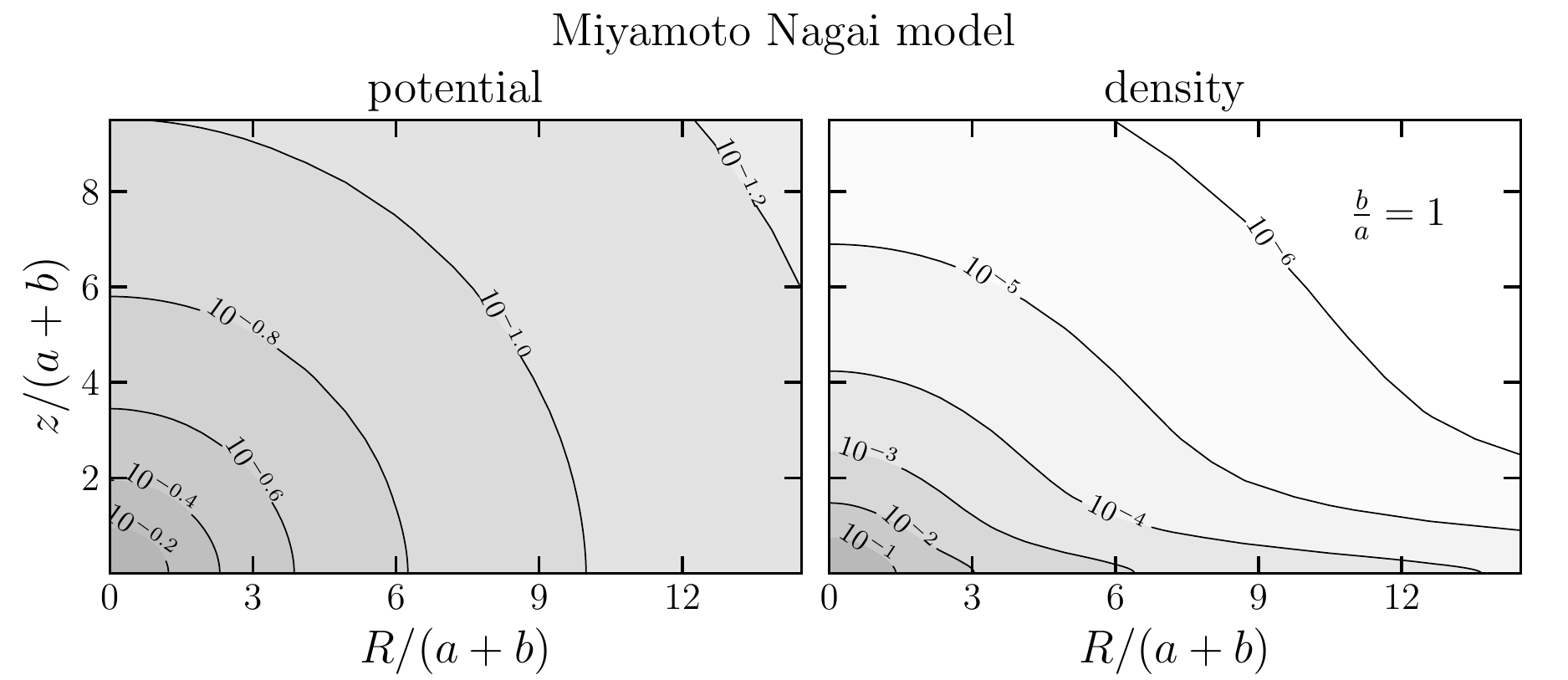}
    \caption{Left panel: iso-potential contours in the meridional plane of a MN model with $a=b$, labelled by the values of $\Phi/\Phi_0$, where $\Phi_0$ is the central potential.
    Right panel: iso-density contours in the meridional plane of the same MN model as in the left panel, labelled by the values of $\rho/\rho_0$, where $\rho_0$ is the central density.}\label{fig:MN}
\end{figure*}

\begin{figure*}
    \centering
    \includegraphics[width=1\hsize]{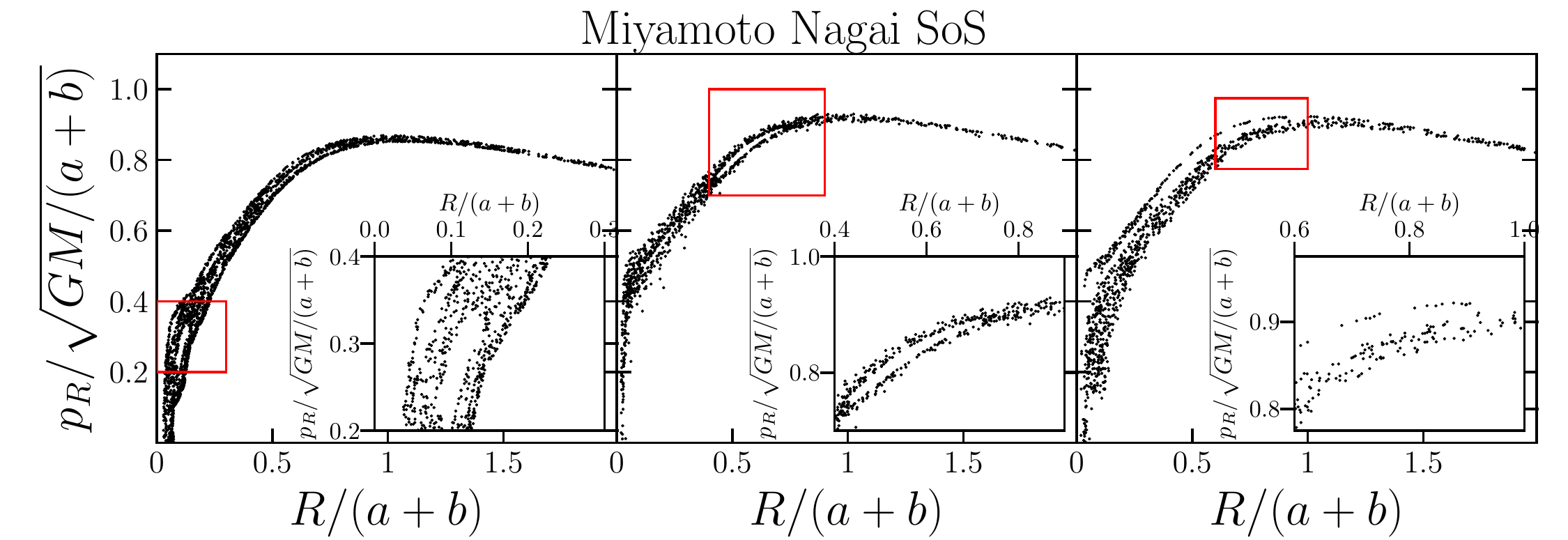}
    \caption{Traces in the SoS of three chaotic orbits in a MN potential with $a=b$. The initial conditions of these orbits (MN1, MN2 and MN3, from left to right) are given in Table~\ref{tab:MNics}. The insets show zoomed-in views of the red boxes in the corresponding main panels.}\label{fig:MNch1}
\end{figure*}

\begin{table*}
    \centering
    \begin{tabular}{ccccccc}
    \hline\hline
    orbit & $R/(a+b)$ & $z/(a+b)$ & $\pR/\sqrt{GM/(a+b)}$ & $\pz/\sqrt{GM/(a+b)}$ & $E/[GM/(a+b)]$ &    $\Lz/\sqrt{GM(a+b)}$     \\
    \hline\hline
    MN1 &   0.05 &	0               &    0          &	-1.3629	    &   -0.05   &   0.01    \\
    MN2 &   47.5482	&   -85.3202    &   -0.011949	&   0.0155498	&   -0.01   &   0.01    \\
    MN3 &   12.2099 &	-31.158	    &	-0.0767776  &   0.181772	&	-0.01   &   0.01    \\
    \hline\hline

    \end{tabular}
    \caption{Same as Table \ref{tab:icSP}, but for the chaotic orbits shown in Fig.~\ref{fig:MNch1}, integrated in the MN potential (equation \ref{for:MNpot}) with $a=b$.}\label{tab:MNics}
\end{table*}


The bottom panel of Fig.~\ref{fig:SPsos} shows in a single SoS the traces of the three orbits in the top panels. The trace of the resonantly trapped orbit is located within the resonant islands of the chaotic orbit (insets in the bottom panel), while the trace of the regular orbit lies outside the region of the SoS occupied by the chaotic orbit. The boundary of resonant islands, representing  the separation between resonantly trapped and chaotic orbits, are believed to correspond to the transition between a region of phase space influenced by only one resonance (within the island) and another region in which more than one resonance is important (outside the island). In the latter case, the star is scattered with no regularity, jumping from a resonance to another (a phenomenon known as resonance overlap; \citealt{Chirikov1979}), which makes the orbit chaotic.  

Having found chaotic orbits, we have demonstrated that, at least for $a/b=0.5$, the shifted Plummer potential is non-integrable. More generally, this is also a proof that
the complexification of a spherically symmetric model does not necessarily produce an integrable axisymmetric potential.

\subsection{Orbits in a Miyamoto-Nagai potential}
\label{sec:mn}

Let us consider the MN potential
\begin{equation}\label{for:MNpot}
 \Phi(R,z)=-\frac{GM}{\sqrt{R^2 + \left[a + \sqrt{b^2 + z^2}\right]^2}},
\end{equation}
where $a$ and $b$ are, respectively, the model's scale radius and scale height, and $M$ is the total mass of the system. The potential~(\ref{for:MNpot}) is typically used to describe the disc components of spiral galaxies since it produces the disk-like density distribution 
\begin{equation}\label{for:MNrho}
 \rho(R,z) = \frac{b^2M}{4\pi}\frac{aR^2+\left[a+3\sqrt{z^2+b^2}\right]\left[a+\sqrt{z^2+b^2}\right]^2}{\left[R^2+\left(a+\sqrt{z^2+b^2}\right)^2\right]^{\frac{5}{2}}\left(z^2+b^2\right)^{\frac{3}{2}}}.
\end{equation}
As shown by \cite{AnEvans2019}, given a spherical potential $\Phi(r)$, the MN substitution that maps
\begin{equation}
 r\to\sqrt{R^2 + \left[a + \sqrt{b^2 + z^2}\right]^2}   
\end{equation}
produces oblate models with analytic density-potential pairs (see also \citealt{Nagai1976,Satoh1980}). In this sense, the potential~(\ref{for:MNpot}) is the generalization of the point mass potential. 
The MN potential (\ref{for:MNpot}) is of particular interest since its limiting cases for $b=0$ and $a=0$ are both integrable. When $b=0$, equation (\ref{for:MNpot}) reduces to the Kuzmin disc \citep{Kuzmin1956,Toomre1963}, known to be of the \Stackel form. When $a=0$ the MN potential becomes the classical Plummer sphere.


As in \cite{Hunter2005}, we study the orbits in a MN model with $a=b$, whose  potential and density maps in the meridional plane are shown in Fig.\ \ref{fig:MN}.
The case considered is a significantly, but not highly flattened oblate model. While \cite{Hunter2005} in his exploration did not  find chaotic orbits in the $a=b$ MN potential, we did find chaotic orbits in the very same potential. A selection of three of these orbits, whose initial conditions are given in Table ~\ref{tab:MNics}, is shown in Fig.~\ref{fig:MNch1}.
The orbit in the left panel has $|\Lz|/\Lcirc(E)=4.57\times10^{-3}$, and the orbits in the middle and right panels have $|\Lz|/\Lcirc(E)=4.26\times10^{-3}$, where $E$ is the orbit's energy and $\Lcirc(E)$ is the magnitude of the angular momentum of a circular orbit in the equatorial plane with $R=3(a+b)$ having energy $E$. Given that our orbits have approximately the same energy of the ones explored by \cite{Hunter2005}, we speculate that the reason why he did not find chaotic orbits is that his orbits have relatively high $\Lz$ ($|\Lz|/\Lcirc=0.1$), while ours have $|\Lz|/\Lcirc\sim 10^{-3}$ (see also Section~\ref{sec:casestudy}).

\section{Mass contributions of chaotic and regular orbits to stellar systems}
\label{sec:chaos} 

The results presented in Section~\ref{sec:orbclass} confirm and strengthen the general finding that, as a rule, non-\Stackel axisymmetric galactic gravitational potentials are non-integrable. Once ascertained the presence of chaotic orbits, the next step is to estimate how much they contribute to a given stellar system. In Section~\ref{sec:method}, we define the fractional mass contributions of chaotic and regular orbits to a stellar system of given DF, and we describe a statistical method to infer the expectation values of these quantities and the related uncertainties from a sample of orbits. Among regular orbits we estimate separately the contributions of resonantly trapped and untrapped orbits. In  Section~\ref{sec:casestudy} we present the results of a case study.

\subsection{Definitions and estimates of the fractional mass contributions of orbit families}
\label{sec:method}

Let us consider a gravitational potential $\Phi(\bbx)$ and a DF $\fxv$ of a tracer population confined by $\Phi$, with finite total mass
\begin{equation}
M=\int f (\bbx,\bbv)\dd^3\bbx\dd^3\bbv,
\end{equation}  
where the integral is extended to the entire phase space. Called $\Vc$ the volume of phase space filled by chaotic orbits, the fractional contribution of chaotic orbits to the total mass is
\begin{equation}
\label{eq:xichaos}
\xic \equiv \frac{\Mc}{M},
\end{equation}
where
\begin{equation}\label{eq:mchaos}
\Mc=\int_{\Vc}f(\bbx,\bbv)\dd^3\bbx\dd^3\bbv
\end{equation}
is the mass contributed by chaotic orbits.
Clearly, the fraction $\xic$ depends both on the gravitational potential and on the tracers' DF. A special case is the one in which the system is self gravitating, so that $\nabla^2\Phi=4\pi G\rho$, where
\begin{equation}
 \rho(\bbx)=\int f\dd^3\bbv.
\end{equation}
In the latter case,  $\xic$ is unique for given $f$, but not for given $\rho$. Two self-gravitating systems can have the same $\rho$ (and thus the same $\Phi$), but different DFs: $\xic$ is in general different for each of these DFs. $\xic$ can be estimated by extracting $N$ orbits from $f$ and counting how many of these $N$ orbits turn out to be chaotic based on any orbit classification method. 

Similarly to $\xic$, we define  the fractional mass contribution of resonantly trapped orbits $\xit\equiv\Mt/M$ with $\Mt=\int_{\Vt} f\dd^3\bbx\dd^3\bbv$, where  $\Vt$ is the phase-space volume occupied by resonantly trapped orbits, and the fractional mass contribution of untrapped orbits $\xiu=1-\xic-\xit$. The fractional mass contribution of regular orbits (including both resonantly trapped and untrapped orbits) is $\xir=\xit+\xiu$.


In practice, to measure $\xic$, $\xit$ and $\xiu$ (and thus $\xir$), we proceed as follows. 
For given $f$ and $\Phi$, we extract $N$ orbits from $f$, i.e.\ $N$ sextuplets of phase-space coordinates ($\bbx$,$\bbv$) drawn from $f$. We integrate in time in the potential $\Phi$ the $N$ orbits and classify them, finding $\Nc$ chaotic orbits, $\Nt$ regular resonantly trapped orbits and $\Nu=N-\Nc-\Nt$ regular untrapped orbits. Straightforward estimates of $\xic$, $\xit$ and $\xiu$ would be $\xic=\Nc/N$, $\xit=\Nt/N$ and $\xiu=\Nu/N$, but these numbers, though giving a rough measure of the fractional contributions of the different families of orbits, are not enough to describe in a statistically meaningful way the results of the numerical experiments, if we do not have a measure of the associated uncertainties. For instance, when $\Nc=0$, one would like to estimate an upper limit on $\xic$, which is expected to be more stringent for larger $N$.


We thus estimate $\xic$, $\xit$ and $\xiu$, and the corresponding uncertainties $\delta\xic$, $\delta\xit$ and $\delta\xiu$ using a Bayesian approach. The joint distribution of $(\Nc,\Nt,\Nu)$ is multinomial with parameters $(N,\xic,\xit,\xiu)$. We model the parameters $(\xic,\xit,\xiu)$ with a symmetric Dirichlet distribution with parameters $(\alpha,\alpha,\alpha)$, which implies that the marginal prior distribution of each component of $(\xic,\xit,\xiu)$ is a beta with parameters $(\alpha, 2\alpha)$, corresponding to prior expected values $\xic=1/3$, $\xit=1/3$ and $\xiu=1/3$, which reflects our prior ignorance. Thanks to the conjugacy of the Dirichlet prior to the multinomial model, the posterior distribution of $(\xic,\xit,\xiu)$ is again Dirichlet with updated parameters $(\alpha+\Nc,\alpha+\Nt,\alpha+\Nu)$ \citep[see, e.g.,][]{Robert2007}. A point estimator of $(\xic,\xit,\xiu)$ is given by the mean of the posterior distribution, that is
\begin{equation}
(\widehat{\xic},\widehat{\xit},\widehat{\xiu})=\left(\frac{\alpha+\Nc}{3\alpha+N},\frac{\alpha+\Nt}{3\alpha+N},\frac{\alpha+\Nu}{3\alpha+N}\right).
\label{for:med}
\end{equation}
The uncertainty associated to posterior estimates is quantified by considering the standard deviation of the marginal posterior distribution, given for $\xic$ by
\begin{equation}
\delta \xic=\sqrt{\frac{(\alpha+\Nc)(2\alpha+N-\Nc)}{(3\alpha+N)^2(3\alpha+N+1)}},
\label{for:std}
\end{equation}
and similarly for $\xit$ and $\xiu$. 

Given that the fractional mass contribution of regular orbits is $\xir=1-\xic$ and that the marginal posterior distribution of $\xic$ is a beta distribution $B(\alpha+\Nc,2\alpha+\Nr)$, where $\Nr=\Nu+\Nt$ is the number of regular orbits, the marginal posterior distribution of  $\xir$ is a beta distribution $B(2\alpha+\Nr,\alpha+\Nc)$.
We take as point estimator of $\xir$ the mean of the marginal posterior distribution
\begin{equation}
\widehat{\xir}=\frac{2\alpha+\Nr}{3\alpha+N},
\label{eq:hatxir}
\end{equation}
and as uncertainty on $\xir$ the standard deviation of the marginal posterior distribution,
\begin{equation}
\delta \xir=\sqrt{\frac{(2\alpha+\Nr)(\alpha+\Nc)}{(3\alpha+N)^2(3\alpha+N+1)}}.
\label{eq:deltaxir}
\end{equation}
In what follows we set $\alpha=1/3$ which, if $N\gg 1$, corresponds to assigning little weight to the prior component when computing the posterior distribution. 

To summarize, in order to quantify the fractional mass contributions of chaotic ($\xic$) and resonantly trapped ($\xit$) orbits to a non-spherical stellar system, either self-gravitating or immersed in an external gravitational potential, one should:
\begin{itemize}
    \item[i)] know (analytically or numerically) the DF of the stellar system, because
    $\xic$ and $\xit$ depend on the DF, and not only on the total gravitational potential and the stellar density distribution;
    \item[ii)] extract from the DF a sample of $N$ phase-space coordinates to be used as initial conditions for $N$ orbits of the stellar system;
    \item[iii)] integrate the orbits in the total gravitational potential and classify them as untrapped regular, trapped regular or chaotic, by means of any suitable classification method (for instance, inspection of the SoS or spectral analysis);
    \item[iv)] use Bayesian estimators such as the ones provided by equations (\ref{for:med}) and (\ref{for:std}) to infer the expectation values of $\xic$ and $\xit$ and the associated confidence intervals.
\end{itemize}


\subsection{A case study: a stellar system with ergodic DF confined by the flattened logarithmic potential}
\label{sec:casestudy}

We now present examples of measures of the fractional mass contributions defined in Section~\ref{sec:method} for specific axisymmetric stellar systems. A natural choice could be to consider self-gravitating stellar systems with either the shifted Plummer or the MN gravitational potentials studied in Section~\ref{sec:orbclass}. However, such systems, though having analytic density distributions (equations~\ref{for:shiftPlumdens} and \ref{for:MNrho}), as far as we know do not have easily tractable analytic DFs\footnote{See \citet{Dejonghe1986} for the analytic expression of the two-integral DF of the self-gravitating MN model.}. Though it is in principle possible to compute and use numerical DFs
\citep[e.g.][]{LyndenBell1962,HunterQian1993,PetacUllio2019}, here we prefer to avoid such a complication and focus on an exceptionally fortunate case in which density, potential and DF have simple analytic expressions.

The stellar system here considered consists of a tracer population with DF $f$ in an external potential $\Phi$. As $f$ we take the ergodic DF of \cite{Evans1993}
\begin{equation}\label{for:EvansDF}
 \fxv = \fnorm \exp{\biggl[-\biggl(\frac{pE(\bbx,\bbv)}{\sigma^2}\biggr)\biggr]},
\end{equation}
where $E\equiv\Phi+K$ is the specific (i.e.\ per mass unit) energy, $K$ is the specific kinetic energy and
\begin{equation}\label{for:DFnorm}
\fnorm = \biggl(\frac{p}{2\pi^2\Rc^2\sigma^2}\biggr)^\frac{3}{2}\frac{M}{q}\frac{\Gamma\bigr(\frac{p}{2}\bigl)}{\Gamma\bigl(\frac{p}{2}-\frac{3}{2}\bigr)}
\end{equation}
normalises the DF to the total mass $M$. The external potential is the flattened logarithmic potential \citep{Binney1981}
\begin{equation}\label{for:pot}
 \Phi(R,z)=\frac{\sigma^2}{2}\ln\left(1 + \Rtilde^2 + \frac{\ztilde^2}{q^2}\right),
\end{equation}
where $\Rtilde\equiv R/\Rc$, $\ztilde\equiv z/\Rc$, $q\le 1$ is the minor-to-major axis ratio of the iso-potential surfaces, $\sigma$ is the maximum circular speed, and $\Rc$ is the core radius (i.e.\ the distance from the center within which the potential is roughly constant). We note that $\Phi(0,0)=0$, so $E\geq0$ for all $({\bf x},{\bf v})$.
The density distribution that generates the gravitational potential is everywhere positive when  $q>1/\sqrt{2}$.
Since the DF (\ref{for:EvansDF}) is ergodic, the density of the tracers stratifies on the iso-potential surfaces of (\ref{for:pot}). 

Integrating the DF~(\ref{for:EvansDF}) over velocities, one finds that the tracers' spatial distribution is 
\begin{equation}\label{for:rho}
 \rho(R,z) = \frac{\rhonorm}{(1 + \Rtilde^2 + \ztilde^2/q^2)^{p/2}},
\end{equation}
with
\begin{equation}\label{for:rhonorm}
 \rhonorm =\frac{M}{\pi^\frac{3}{2}\Rc^3} \frac{\Gamma(\frac{p}{2})}{q\Gamma(\frac{p}{2}-\frac{3}{2})} 
 \end{equation}
\citep{Evans1993}. 
The tracers' density distribution (\ref{for:rho}) is a power-law of slope $p$ at large radii, while in the central regions a core of approximately constant density extends out of $\Rc$. Since the DF is ergodic, the tracers' velocity distribution is isotropic and the second velocity moments are
\begin{equation}
\overline{\vR^2} = \overline{\vphi^2} = \overline{\vz^2} = \frac{\sigma^2}{p}.
\end{equation}


\begin{figure*}
    \centering
    \includegraphics[width=.9\hsize]{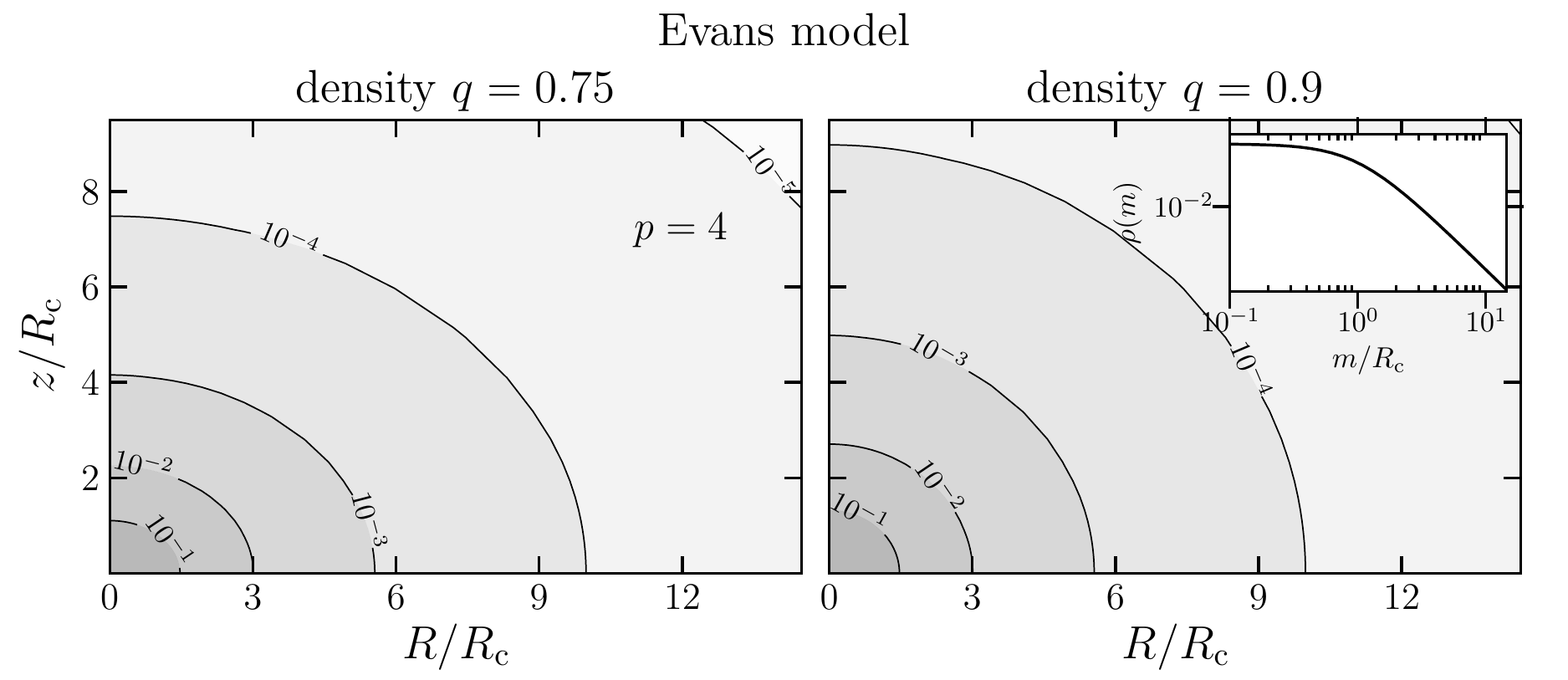}
    \caption{Left panel: iso-density contours in the meridional plane of the density distribution~(\ref{for:rho}) for $p=4$ and $q=0.75$, labelled by the values of $\rho/\rho_0$, where $\rho_0$ is the central density.  Right panel: same as the left panel but for $q=0.9$.
    The inset in the right panel shows the density of the models as a function of the elliptical radius $m\equiv\sqrt{R^2+z^2/q^2}$.}\label{fig:Evans}
\end{figure*}

We focus on two specific models with $p=4$, which differ only in the value of the potential's axis ratio $q$: a highly flattened model with $q=0.75$, close to the minimum $q$ allowed for consistency, and an almost spherical model with $q=0.9$. Fig.~\ref{fig:Evans} shows the isodensity contours of the $q=0.75$ (left-hand panel) and $q=0.9$ (right-hand panel) models. The small inset in the right-hand panel shows the density as a function of the elliptical radius $m\equiv\sqrt{R^2+z^2/q^2}$, which is the same for both models (equation~\ref{for:rho}), with $\rho\propto m^{-4}$ when $m\gg \Rc$. $\Rc$ and $\sigma$ set the physical scales of the model. 

For each model we sample a set of $N=10000$ orbit initial conditions from the DF (\ref{for:EvansDF}), following the procedure\footnote{This procedure does not exploit the fact that the DF is ergodic and can be used with any DF $f(\bbx,\bbv)$.} described in Appendix~\ref{sec:samp} . Each orbit is integrated in the potential~(\ref{for:pot}) for $\approx 10^4\too$, with $\too\equiv\Rc/\sigma$, and the SoS trace of each orbit is computed.
For $p=4$ the model's circularized half-mass radius is $\rh\simeq2.2\Rc$, so 
$\too\simeq\rh/(2.2\sigma)$. When the system is scaled to represent a typical massive elliptical galaxy
with effective radius $\Reff\approx 10\kpc$ and central stellar velocity dispersion $\sigma_0\approx 250\kms$ \citep*[e.g.\ section 5.4 of][]{CFN19}, assuming $\rh\approx\Reff$ and $\sigma\approx\sigma_0$, we get $\too\approx 18\Myr$. In all cases, the traces in the SoS are sampled with at least 1000 points. 


\begin{figure*}
 \centering
 \includegraphics[width=1.\hsize]{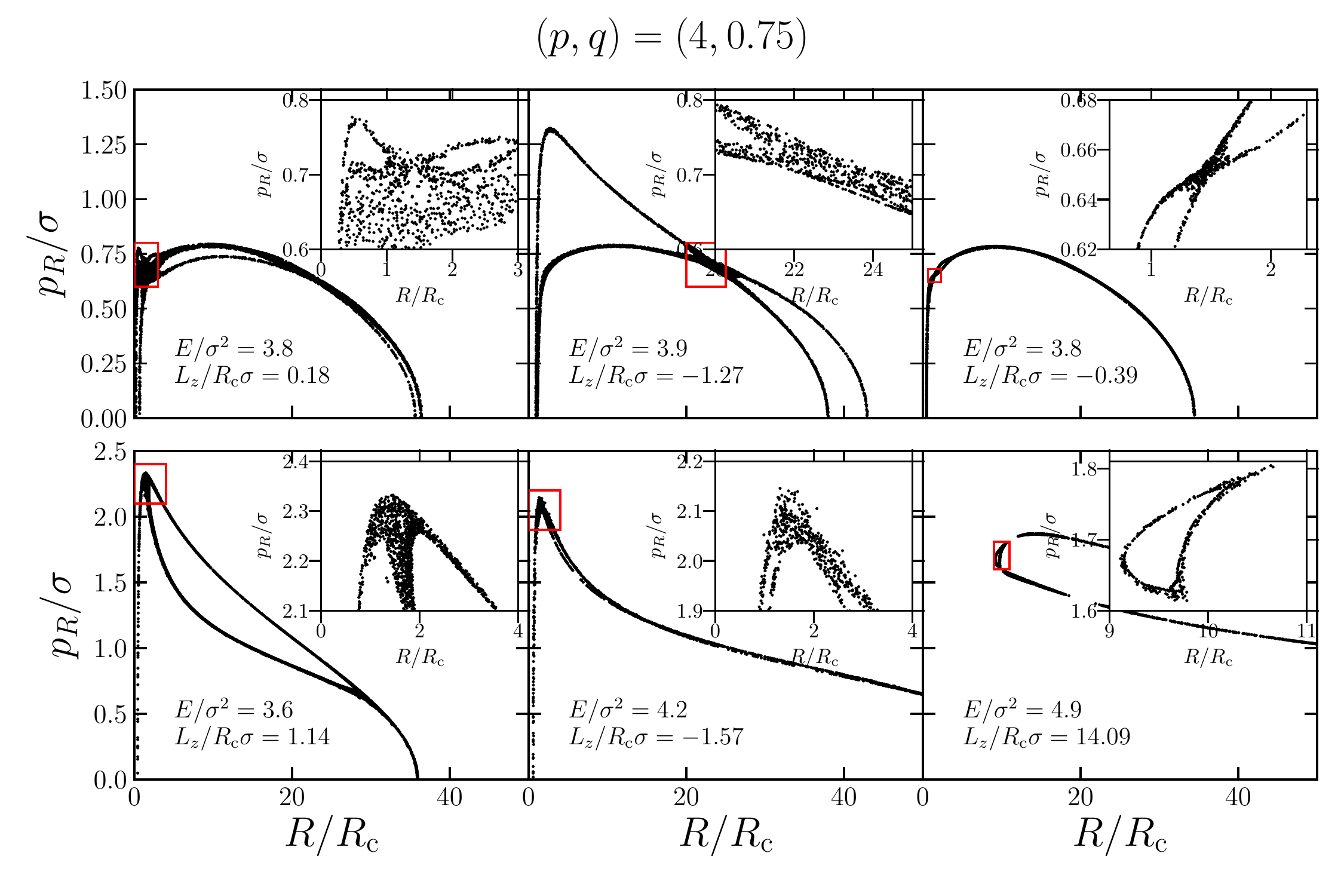}
 \caption{Traces in the SOS of a selection of six of the chaotic orbits found for the stellar system with gravitational potential (\ref{for:pot}) with $q=0.75$.  The inset in each panel shows a zoomed-in view of the region marked with a red box in the corresponding SoS.}\label{fig:chaos}
\end{figure*}

\begin{table}
\begin{center}
 \caption{Number of orbits, out of $N=10000$, classified as chaotic ($\Nc$), regular ($\Nr=\Nu+\Nt$), untrapped ($\Nu$) and resonantly trapped ($\Nt$), and corresponding fractional mass contribution of each class of orbits ($\xic$, $\xir$, $\xiu$, $\xit$) for a more flattened ($q=0.75$) and a less flattened ($q=0.9$) axisymmetric stellar system with ergodic Evans DF (\ref{for:EvansDF}) with $p=4$ and potential~(\ref{for:pot}). The value of $\xi_i$, for $i=$c, r, u and t, is given in the form $\widehat{\xi}_i\pm\delta\xi_i$, where $\widehat{\xi}_i$ and $\delta\xi_i$ are, respectively, the mean and the standard deviation of the posterior distribution of $\xi_i$.}
\begin{tabular}{lrr}
 \hline\hline
	&		  $q=0.75$ 	& 	$q=0.9$		\\
 \hline\hline
$\Nc$             & $20$              &  $5$ \\
$\Nr=\Nu+\Nt$		& $9980$		& $9995$	\\ 
$\Nu$		& $9419$		& $9859$	\\ 
$\Nt$		& $561$		& $136$	\\ 
$\xic$        & $(2.03\pm0.45)\times 10^{-3}$ &  $(5.33\pm2.31)\times 10^{-4}$\\
$\xir=\xit+\xiu$		& $0.9980\pm0.0012$		& $0.99947\pm 0.00023$	\\
$\xiu$		& $0.9418\pm0.0023$		& $0.9858\pm 0.0012$	\\
$\xit$        & $(5.61\pm0.23)\times 10^{-2}$ &  $(1.36\pm0.12)\times 10^{-2}$\\
 \hline\hline
\end{tabular}\label{tab:xi}
\end{center}
\end{table}

For both $q=0.75$ and $q=0.9$ we find, among the explored orbits, a few chaotic orbits, which demonstrates that logarithmic potentials with these flattening parameters are non-integrable. Fig.~\ref{fig:chaos} shows the SoS of a selection of six of the orbits classified  as chaotic in the $q=0.75$ model. To highlight the two-dimensional structure of the manifold on which the consequents lie, the small insets in each panel show a zoom-in of a portion of the corresponding SoS. While we do find chaotic orbits, \cite{Richstone1982} and \cite{Barnes2001} did not find any chaotic orbit in the same logarithmic potential with $q=0.75$, probably due to the low statistics of their sample of orbits. \cite{Richstone1982} explored only 400 orbits, when, for instance, our results show that on average only one out of about 500 orbits is chaotic if the phase-space  is populated with the DF (\ref{for:EvansDF}) with $p=4$. \citet{Barnes2001} did explore even fewer orbits than \cite{Richstone1982}.

The number of chaotic ($\Nc$), regular ($\Nr=\Nt+\Nu$), resonantly trapped ($\Nt$) and untrapped ($\Nu$) orbits found for both $q=0.75$ and $q=0.9$ are reported in Table~\ref{tab:xi}, together with estimates of the corresponding fractional mass contributions $\xic$, $\xir=\xit+\xiu$, $\xit$ and $\xiu$, obtained from equations~(\ref{for:med}-\ref{eq:deltaxir}) with $\alpha=1/3$.
While in both cases the regular orbits are by far the dominant family ($\xir\gtrsim 99.8\%$), the much rarer chaotic orbits contribute more to the more flattened system ($\xic\simeq 0.002$ for $q=0.75$, and $\xic\simeq 0.0005$ for $q=0.9$).  The considered systems thus appear largely regular, with small contributions from chaotic orbits. For these chaotic orbits our analysis (based on SoS traces obtained with long time integration) does not provide information on the characteristic chaotic timescale, that is the time over which the orbits starts showing a chaotic behaviour. It is then possible that a fraction of the found chaotic orbits are sticky \citep[e.g.][]{Maffione2015}, i.e.\ that they behave similarly to regular orbits for relatively long time, before manifesting their chaotic nature. If this is the case, we would
have found even smaller values of $\xic$ if we had integrated the orbits for shorter times, more realistic for astrophysical applications. However,
the adopted SoS orbit classification method forces us to consider long integration times, because a reliable classification requires that the orbit's trace in the SoS has a large number of consequents.

The mass contribution of resonantly trapped orbits is almost a factor of 30 higher than that of the chaotic orbits for both $q=0.75$ and $q=0.9$: in particular, the overall contribution is non-negligible in the case of the more flattened model $(q=0.75)$, in which about $6\%$ of the stellar mass is in resonantly trapped orbit. This is qualitatively\footnote{Richstone's sample of orbits is not extracted from a DF, so the comparison is not quantitative.} consistent with the results of  \cite{Richstone1982} who found that, in the same $q=0.75$ logarithmic potential, $\approx 5\%$ of the orbits of his sample are resonantly trapped (note that Richstone adopts a different nomenclature in which the resonantly trapped orbits are called pipe orbits).

\begin{figure*}
    \centering
    \includegraphics[width=.9\hsize]{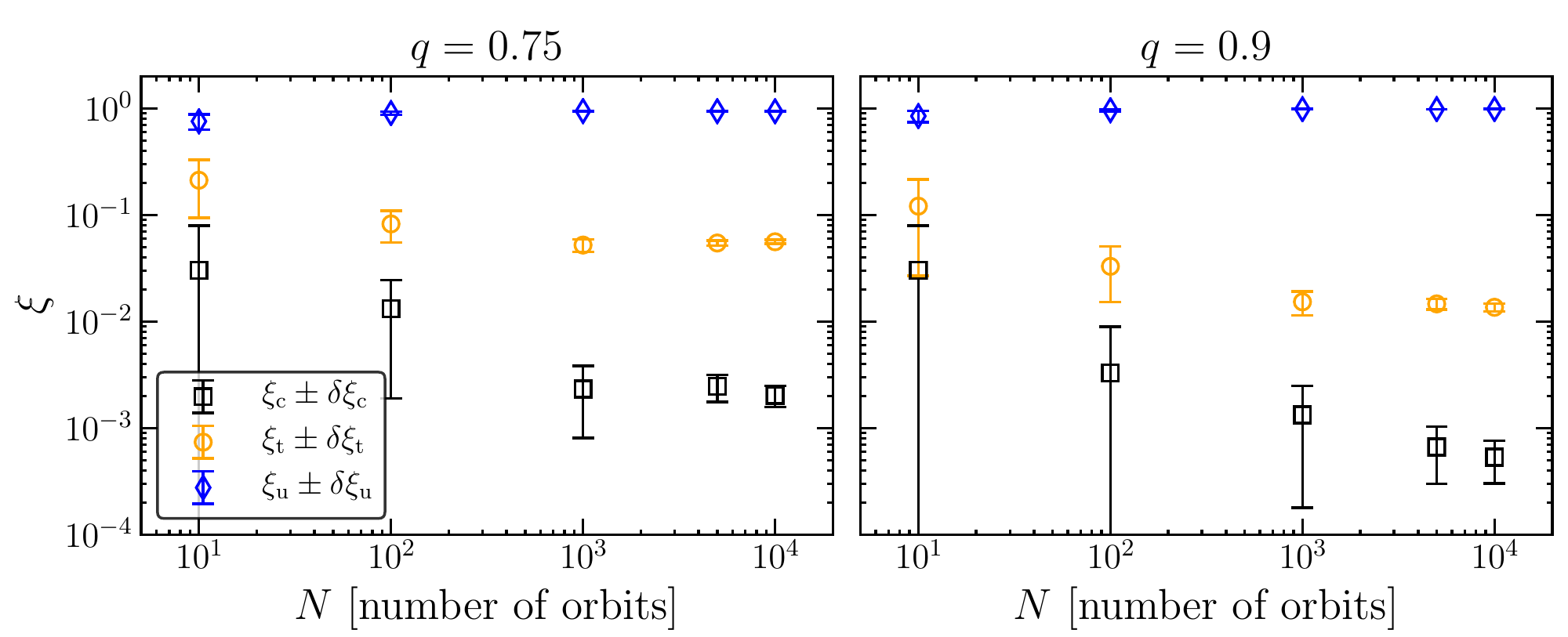}
    \caption{Estimates of the fractional mass contributions of chaotic ($\xic$; black squares with errorbars), trapped ($\xit$; orange circles with errorbars) and untrapped ($\xiu$; blue diamonds with errorbars) orbits obtained by selecting random subsamples of $N$ orbits among our samples of 10000 orbits for the $q=0.75$ (left panel) and $q=0.9$ (right panel) Evans models.}
    \label{fig:xi}
\end{figure*}

We note that we are able to provide estimates of $\xic$, $\xit$, $\xiu$ and $\xir$ with relatively small associated uncertainties (see Table~\ref{tab:xi}), which means that the number of explored orbits ($N=10000$ for each stellar system) is sufficient for our purposes. The estimates of the fractional mass contributions deteriorate with decreasing $N$. This is illustrated quantitatively by Fig.~\ref{fig:xi}, showing, for both $q=0.75$ and $q=0.9$, the estimates of $\xic$, $\xit$ and $\xiu$ as functions of $N$, that we obtained by selecting random subsamples of $N$ orbits among our samples of 10000 orbits.
As expected, the estimates for different $N$ are statistically consistent, but the error bars on the fractional mass contributions shrink monotonically for increasing $N$.

\begin{figure}
 \centerline{ \includegraphics[width=\hsize]{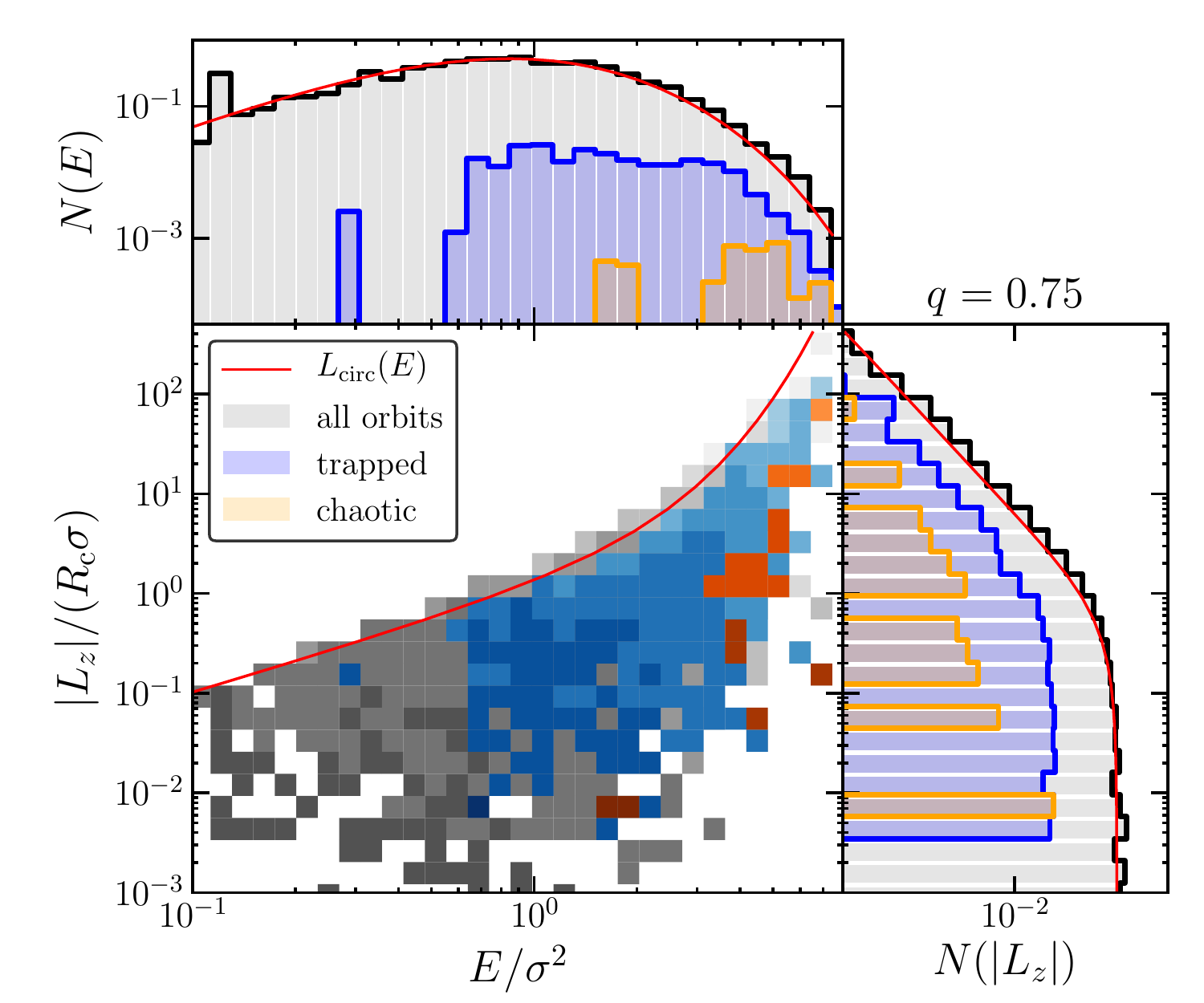}}
 \caption{Distributions in the space of the orbital parameters $E$ and $\Lz$ for our sample of orbits integrated in the flattened logarithmic potential (\ref{for:pot}) with $q=0.75$. {\em Lower-left panel}: number density distribution of the explored orbits in the $(E,|\Lz|)$ plane, for all orbits (grey), resonantly trapped orbits (blue), and chaotic orbits (orange): the darker the pixel's colour the higher the number density. The red curve indicates $|\Lcirc(E)|$ for circular orbits in the equatorial plane. {\em Upper panel}. Differential energy distribution $N(E)$ for all orbits (grey), resonantly trapped orbits (blue) and chaotic orbits (orange), for the same sample of orbits as in the lower-left panel. The red curve indicates $N(E)$ as computed from the DF from which the sample of orbits is extracted (see Appendix~\ref{app:C}). {\em Right panel}. Differential $\Lz$ distribution  $N(|\Lz|)$ for all orbits (grey), resonantly trapped orbits (blue) and chaotic orbits (orange), for the same sample of orbits as in the lower-left panel. The red curve indicates $N(|\Lz|)$ as computed from the DF from which the sample of orbits is extracted (see Appendix~\ref{app:C}).}
\label{fig:ne_nlz_q075}
\end{figure}
\begin{figure}
 \centerline{
 \includegraphics[width=\hsize]{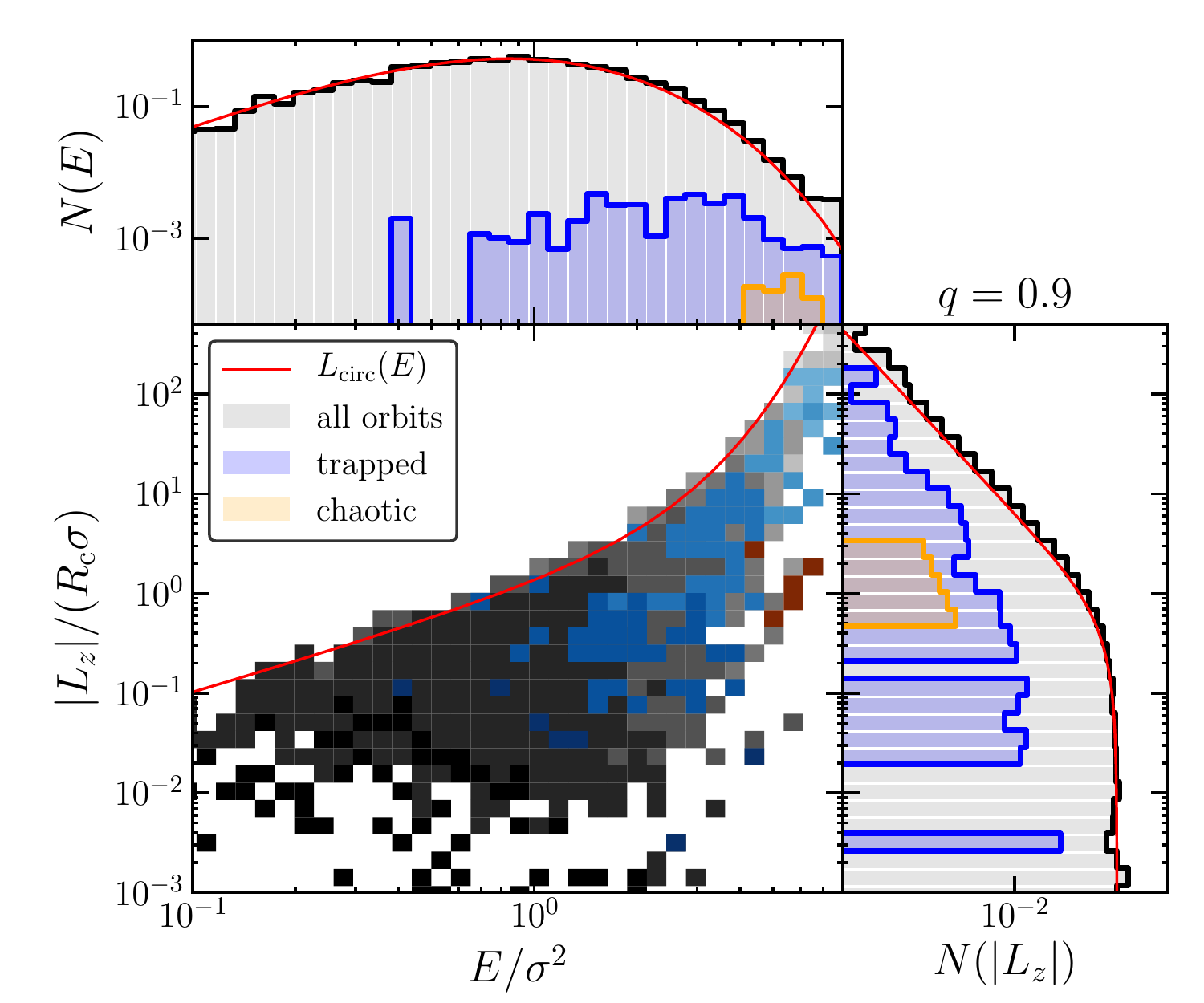}}
 \caption{Same as Fig. \ref{fig:ne_nlz_q075}, but for  the sample of orbits integrated in the flattened logarithmic potential with $q=0.9$.}
\label{fig:ne_nlz_q09}
\end{figure}

In Fig.s~\ref{fig:ne_nlz_q075} and \ref{fig:ne_nlz_q09} we show the two-dimensional, joint distributions $n(E,|\Lz|)$ for $q=0.75$ and $q=0.9$, respectively. $n(E,|\Lz|)$ is such that $n(E,|\Lz|)\d E \d |\Lz|$ is the fraction of orbits with energy between $E$ and $E+\d E$ and absolute value of $\Lz$ between $|\Lz|$ and $|\Lz|+\d|\Lz|$. In each figure, the top panel shows the differential energy distribution $N(E)$, while the left panel the differential $\Lz$ distribution $N(|\Lz|)$. $N(E)\d E$ gives the fraction of orbits with energy between $E$ and $E+\d E$. Similarly, $N(|\Lz|)\d|\Lz|$ is the fraction of orbits  with absolute value of $\Lz$ between $|\Lz|$ and $|\Lz|+\d|\Lz|$ (for details, see Appendix~\ref{app:C}). We have marked with different colours the contributions to the one- and two-dimensional distributions of the chaotic and resonantly trapped orbits. We note that as the energy increases, the relative contributions of the resonantly trapped and of the chaotic orbits increase while, especially for small values of $|\Lz|$, their fractional contribution tends to remain approximately constant with $|\Lz|$.

As an example of specific application, for instance within the framework of the study of the stellar streams generated by globular clusters (see Section~\ref{sec:intro}), the Evans density distribution (\ref{for:rho}) can be interpreted as a simple model of the globular cluster system of a galaxy, whose gravitational potential is given by equation~(\ref{for:pot}). Our analysis provides the fraction of these globular clusters that are expected to be on chaotic or trapped orbits, under the assumption that the velocity distribution of the globular cluster system is isotropic. Fig.s~\ref{fig:ne_nlz_q075} and \ref{fig:ne_nlz_q09} suggest that these fractions should be similar for anisotropic velocity distribution, corresponding to DFs that either favour or disfavour high-$|\Lz|$ orbits. The very small values found for $\xic$ in the two explored models indicate that the fraction of GC streams dispersed by chaos would be negligible. The fraction of GC streams dispersed by separatrix divergence is not directly measured by our calculations, but $\xit$ can be taken as an upper limit on this fraction, under the  plausible assumption that there are more orbits belonging to resonantly trapped families than close to their boundaries.


\section{Conclusions}
\label{sec:conc}

We have addressed the problem of the integrability of a few families of axisymmetric potentials. Using numerical orbit integration it is possible to show that a given potential is non-integrable, but of course it is not possible to demonstrate that it is integrable. We have added to the literature a few new cases of axisymmetric potentials that turn out to be non-integrable. These potentials belong to the families of shifted Plummer, MN and flattened logarithmic potentials.

Given that chaotic orbits have been found for all (non-St\"ackel) axisymmetric potentials in which they have been deeply looked for, an interesting question is how much these chaotic orbits contribute to stellar systems that are confined by these potentials. We have defined the fractional mass contribution $\xic$ of chaotic orbits in stellar systems of given DF and proposed a simple but robust statistical method to estimate it from a sample of orbits. With the same approach one can estimate also the fractional mass contributions  of resonantly trapped regular orbits ($\xit$), of untrapped regular orbits ($\xiu$) and of all regular orbits ($\xir=\xit+\xiu$).

As a case study, we analysed two axisymmetric stellar systems with Evans DF confined by flattened logarithmic gravitational potentials with different values of the axis ratio $q$. The contribution of chaotic orbits is extremely small in both cases, but significantly higher in the more flattened ($q=0.75$) system ($\xic = 0.0020 \pm 0.0005$) than in the almost spherical ($q=0.9$) system ($\xic = 0.0005 \pm 0.0002$). In both cases the mass contribution of resonantly trapped orbits is almost a factor of 30 higher than that of the chaotic orbits ($\xit=0.056\pm 0.002$ for $q=0.75$ and $\xit=0.014\pm 0.001$ for $q=0.9$). Most ($\gtrsim 94\%$) of the mass of these  axisymmetric system is contributed by the standard untrapped regular orbits. The presented case study is particularly simple, because the considered axisymmetric stellar systems have analytic gravitational potential $\Phi$, analytic ergodic DF $f$ and analytic spatial density $\rho$ of the tracer population. However, we stress that the same analysis can be performed when the DF is not ergodic and/or one or more among $\Phi$, $f$ and $\rho$ are not known analytically.

A fundamental step of our analysis is classifying orbits as chaotic and regular, and, among regular orbits, distinguishing resonantly trapped and untrapped orbits. In this paper we have used as orbit classification technique the visual inspection of the SoS, but we stress that any technique for orbit classification can be used. Of course, visual inspection is impractical if one wants to systematically classify very large samples of orbits, which requires automatic classification algorithms \citep[e.g.][]{Carpintero2014}. We have seen that, from a geometrical point of view, orbits can be considered lower dimensional manifolds embedded in a higher dimensional space (the full phase space). To distinguish, at low computational cost, chaotic and regular orbits one could estimate the intrinsic dimension of these manifolds \citep{Mordohai2005} and build probabilistic models of a few selected prototypes (i.e.\ using generative topographic mapping; \citealt{Bishop1998}). Alternative methods based on the correlation integrals are also a viable way to estimate the dimensionality of the orbit in phase space \citep{Carnevali1984,Barnes2001,Carpintero2008} or of the trace in the SoS. However, determining the dimensionality of the orbit manifold is not enough to discriminate between untrapped and  resonantly trapped orbits. Such a task requires either spectral methods \citep{BinneySpergel1982,CarpinteroAguilar1998} or an algorithm able to classify topologically the traces in the SoS.

\section*{Acknowledgements}

We are grateful to J.\ Binney, J.\ Magorrian and B.\ Nipoti for helpful discussions. We thank an anonymous referee for comments and suggestions that helped improve this work. RP
acknowledges G.\ Sabatini and G.\ Porrino for useful suggestions, comments and moral support.

\section*{DATA AVAILABILITY}
The data underlying this article will be shared on reasonable request to the corresponding author.

\bibliography{paper}
\bibliographystyle{mnras}

\appendix





\section{Extracting orbits from a distribution function}
\label{sec:samp}

We extract $N$ orbits with phase-space coordinates $\bbw_k\equiv\{\xk,\yk,\zk,\vxk,\vyk,\vzk\}$, with $k=1,...,N$, from a DF $f(\bbx,\bbv)$ using a Metropolis-Hastings \citep{Metropolis1953,Hastings1970} sampler. Here we briefly describe the procedure.

\begin{itemize}
\item[i)] From a given point in the phase space $\bbw_{l}$, a new location $\bbw_{l+1}$ is sampled using as proposal distribution a multivariate Gaussian distribution $\multiN$ centered in $\bbw_l$. The covariance matrix $\bbC^{2}$ of $\multiN$ is given by 
 (e.g. \citealt{Roberts1997, Rosenthal2010})
 \begin{equation}
  \bbC^2 = \biggr(\frac{2.38}{d}\biggl)^2 (\bbsigma^2 + \varepsilon \mathbb{I}),
 \end{equation}
 where $d=6$ is the dimension of the phase space, $\mathbb{I}$ is the identity matrix, $\varepsilon=10^{-8}$ and $\bbsigma^2$ is the chain's empirical covariance matrix, whose $(i,j)$-element is
 \begin{equation}\label{for:covar}
  \sigma^2_{i,j} = \frac{1}{\Mmin}\sum_{m=M-\Mmin}^M 
  (w_{i,m} - \overline{w_i})(w_{j,m} - \overline{w_j}),
 \end{equation}
 with $M$ the current chain size. The empirical covariance matrix and $\overline{w_h}$ (i.e. the mean over the $h$-th parameter of $\bbw$) are constructed using only the chain's latest $\Mmin=10000$ steps, allowing the sampler to adapt to the local structure of the probability distribution, and a new covariance matrix is built every 1000 steps. The term $\varepsilon\mathbb{I}$ avoids $\bbC^2$ to collapse to zero, especially during the first chain's steps, where the empirical covariance matrix cannot be recovered.

 \item[ii)] At each draw, the quantity $P\equiv \min[1,f(\bbw_{l+1})/f(\bbw_l)]$ is computed. The proposal $\bbw_{l+1}$ is accepted if $P=1$ and, in this case, the next chain step starts from $\bbw_{l+1}$. Otherwise, the new proposal is accepted only if $P>U$, with $U$ drawn from a uniform distribution in the interval $]0,1[$.

 \item[iii)] To build a sample of $N$ elements, we use a chain with $\Nchain$ steps, 
 where $\Nchain=\Nbi+N\delta n$, because we eliminate the first $\Nbi$ burn-in steps, and, of the remaining chain, we take one draw every $\delta n$ iterations. In particular, we adopt $\Nbi=50000$ and $\delta n=200$, which guarantees that the final sample does not contain duplicate elements and that the sample's autocorrelation is sensibly lowered. 
 \end{itemize}

\section{Useful formulae}
\label{app:C}

Here, we report the expressions of the differential energy distribution and of the differential $\Lz$ distribution computed from the DF~(\ref{for:EvansDF}), which are shown in Fig.s~\ref{fig:ne_nlz_q075} and \ref{fig:ne_nlz_q09}.

The differential energy distribution is \citep{BinneyTremaine2008}
\begin{equation}\label{for:fE}
 N(E) = \fEnorm \exp\biggl[-\frac{pE}{\sigma^2}\biggr]g(E;\sigma),
\end{equation}
where $g$ is the non-normalized density of states, defined as
\begin{equation}\label{for:int4}
 g(E;\sigma)=\int_0^{\mmax} m^2\sqrt{\biggl[\frac{2E}{\sigma^2}-\ln(m^2 + 1)\biggr]}\dd m,
\end{equation}
with \begin{equation}
 \mmax=\sqrt{\exp\biggl({\frac{2E}{\sigma^2}}\biggr)-1},
\end{equation}
i.e. the root of the radical in equation (\ref{for:int4}), and 
\begin{equation}
 \fEnorm = \frac{2^{5/2}p^{3/2}}{\sigma^2\pi}\frac{\Gamma\bigr(\frac{p}{2}\bigl)}{\Gamma\bigl(\frac{p}{2}-\frac{3}{2}\bigr)}
\end{equation}
is such that $\int_0^{+\infty} N(E)\dd E=1$.

In analogy with equation~(\ref{for:fE}), we define the differential $\Lz$ distribution (i.e.\ the number of orbits with $|\Lz|$ in the interval [$|\Lz|$,$|\Lz|+\dd|\Lz|$]) as
\begin{equation}\begin{split}\label{for:fLz}
 & N(|\Lz|) = \\
 & \fLznorm\biggl[\int_0^{\infty}\frac{1}{(1+t^2)^{\frac{p-1}{2}}}\exp\biggl(-\frac{p\Lz^2}{2\sigma^2\Rc^2t^2}\biggr)\dd t\biggr]\dd|\Lz|, \\
\end{split}\end{equation}
and
\begin{equation}
 \fLznorm = \biggl(\frac{2p}{\pi}\biggr)^{\frac{1}{2}}\frac{p-3}{\Rc\sigma}
\end{equation}
is such that $\int_0^{\infty} N(|\Lz|)\dd|\Lz|=1$.

\end{document}